\documentclass[graybox]{svmult}

\usepackage{mathptmx}       
\usepackage{graphicx}        

\begin{document}

\title*{Multi-messenger emissions from Kerr black holes}
\author{Maurice H.P.M. van Putten}
\institute{Le Studium IAS, Chair d'Astrophysique, 3D Avenue de la Recherche, 
           45071 Or\'eans, France,
           \email{mvp@ligo.mit.edu}. }
\maketitle

\abstract{Kerr black holes are energetically similar to spinning tops accompanied by frame 
dragging in the surrounding spacetime. Frame dragging is shown herein to be a universal 
causal agent for producing multi-messenger emissions. We discuss high energy emissions produced 
by gravitational spin-orbit coupling along the axis of rotation and low energy emissions from 
surrounding matter via a torus magnetosphere. Model results point to ultra-high energy cosmic rays 
(UHECRs) from supermassive black holes at about the Greisen-Zatsepin-Kuzmin (GZK) energy threshold 
from low-luminosity active galactic nuclei (AGN), and to high-energy photon emissions from stellar
mass black holes in ultra-relativistic capillary jets. The former compares favorably with recent
results by the Pierre Auger Observatory (PAO). The latter compares favorably with gamma-ray
burst data from the High Energy Transient Exporer (HETE) II, {\em Swift} and the Burst and Transient 
Source Experiment (BATSE), wherein a finite lifetime of black hole spin is found to improve the 
correlation between peak energies and true energies in gamma-rays. Matched filtering applied to 600 
light curves of long GRBs identifies a process of viscous spin down against matter at the inner most 
stable circular orbit. We conclude that long GRBs are spin powered, not accretion powered, from 
core-collapse supernovae and merger events such as GRB060614. Contemporaneous low energy emissions
are expected from surrounding matter in gravitational waves by a pressure driven 
Papaloizou-Pringle instability, and more so than in accompanying MeV-neutrinos and magnetic winds. 
The latter powers aspherical radio-loud supernovae in collapsars and long duration extragalactic radio 
bursts from GRB060614 type events. This outlook is of interest to emerging
multi-messenger surveys.}

\section{Introduction}
\label{sec:1}
Recent observations show a Transient Universe abundant in ultra-high energy cosmic rays (UHECRs) \cite{abr07,abr08}, 
gamma-ray bursts (GRBs) \cite{kou93,del06,abd09} and, possibly, extragalactic radio bursts \cite{lor07}. At the 
highest energies, these emissions are non-thermal, generally described by power-law spectra distinct from black body 
radiation. This feature reflects particle acceleration in low-opacity regions, where radiation
can remain out of thermal equilibrium, even as their inner engines appear to be compact as 
inferred from short time scales of variability \cite{pir98}.

Notable candidates for these transient sources are nuclei harboring black holes \cite{lyn69,pac91,woo93,pac98,lev93}.
The discovery of Cygnus X-1 \cite{gia67,bol71,oda71} marks the first candidate black hole of mass 
$M_H\simeq 8.7 M_\odot$ \cite{lor08}. It appears to be in a state of accretion with
Eddington luminosity in the X-rays of less than 0.1 \cite{gil97,zio05,lor08}.
Black holes transient sources offer unique opportunities for probing some non-Newtonian aspects of gravitation with emerging 
multi-messenger surveys of cosmic rays, high-energy photons from X-rays to TeV emissions, neutrino 
emissions and gravitational waves--the emerging field of Astroparticle Physics.

In this invited Lecture, I discuss the physics of radiation around Kerr black holes \cite{ker63} produced by
frame dragging. Kerr black holes are described merely by their mass and angular momentum, and they 
offer unique potential for unification of transient emissions from supermassive in AGN and, from 
stellar mass black holes, in gamma-ray bursts from core-collapse supernovae and 
mergers of black holes and neutron stars, and extragalactic radio bursts.

The flow of this Lecture is as follows. We first formulate some model challenges posed by recent 
observations of UHECRs and long GRBs. We interpret these as radiation phenomena induced by frame dragging 
around Kerr black holes, introducing high-energy emissions by a gravitational spin-orbit coupling along 
the spin axis of the black hole \cite{van00,van05,van08a}. As a gravitational interaction, it is universal and irrespective 
of the nature of the angular momentum of the accelerated particles, whether mechanical or 
electromagnetic in origin. Frame dragging also acts on the torus magnetosphere surrounding the
black hole \cite{van99}. It opens a diversity in radiation channels in catalytic conversion of black hole spin 
energy and angular momentum by surrounding matter in the shape of a hot and magnetized torus \cite{van03}. The 
latter stimulates the black hole to be luminous in a state of suspended accretion \cite{van01a}, wherein 
no net energy or angular momentum is extracted from the torus. The torus forms a boundary layer between the 
black hole and the extended accretion disk. It creates a novel channel for emissions in the radio and, 
when sufficiently hot and slender, in gravitational waves by excitation of non-axisymmetric instabilities
\cite{van01}. In general terms, gravitational waves are important in providing an efficient channel for 
radiating angular momentum, tightly correlated to the evolution of the black hole.

In \S2, we outline some challenges posed by contemporary observations of UHECRs and GRBs. 
In \S3, we summarize some physical properties of Kerr black holes and the energetic interactions due to frame dragging.
In \S4, we introduce some general considerations for modeling black hole radiation processes.
\S5 introduces two frame dragging induced radiation processes. 
In \S6, we apply the first to high-energy emissions from supermassive black holes with recent UHECR 
data by the Pierre Auger Observatory (PAO) and a presently tentative UHECR-AGN association. For stellar 
mass black holes, we compare a model spectral-energy correlation with GRB data from the High Energy 
Transient Explorer (HETE) II and {\em Swift}.
In \S7, we apply the second to derive the evolution of black hole spin in suspended accretion, and a model light curve for
GRB emissions to enable comparison with light curves of long GRBs in the Burst and Transient Source Experiment 
(BATSE) by application of matched filtering. The detailed results of matched filtering point to a dominance
in gravitational wave emissions, as discussed in \S8. The {\em Swift} discovery of X-ray tails is
discussed in \S9, and \S10 comments in the discovery of delayed high energy emissions in GRB080916C
\cite{taj09}. We conclude with an outlook on new multi-messenger transient sources of interest to 
the emerging surveys in the optical-radio and gravitational waves in \S11.

\section{Challenges in high-energy emissions in UHECRs and GRBs}

High energy emissions in cosmic rays and gamma-ray bursts are characteristically non-thermal, 
and are generally well fit by a combination of power-laws.
 
\begin{figure}
\centerline{
\includegraphics[scale=.32]{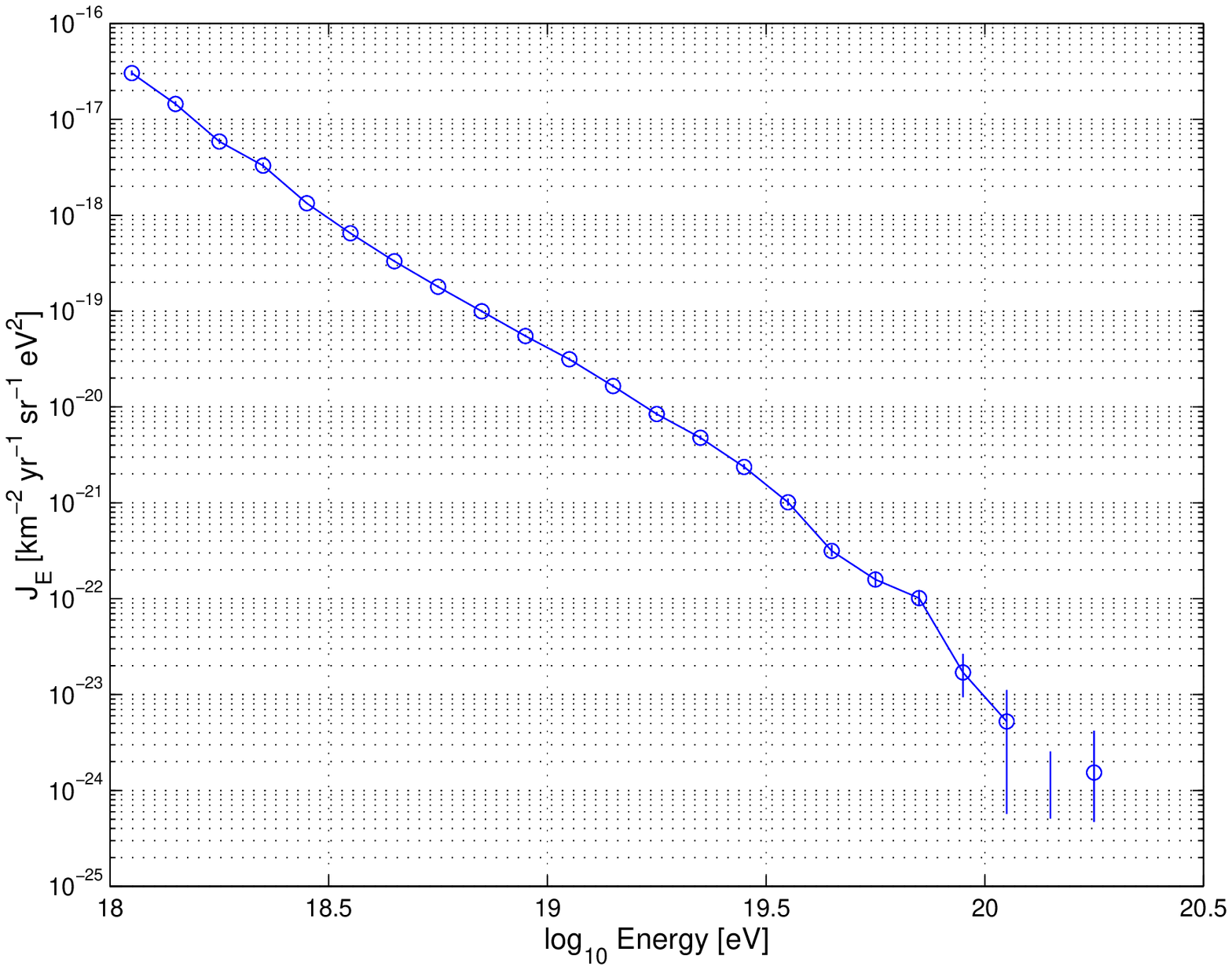}\includegraphics[scale=0.32]{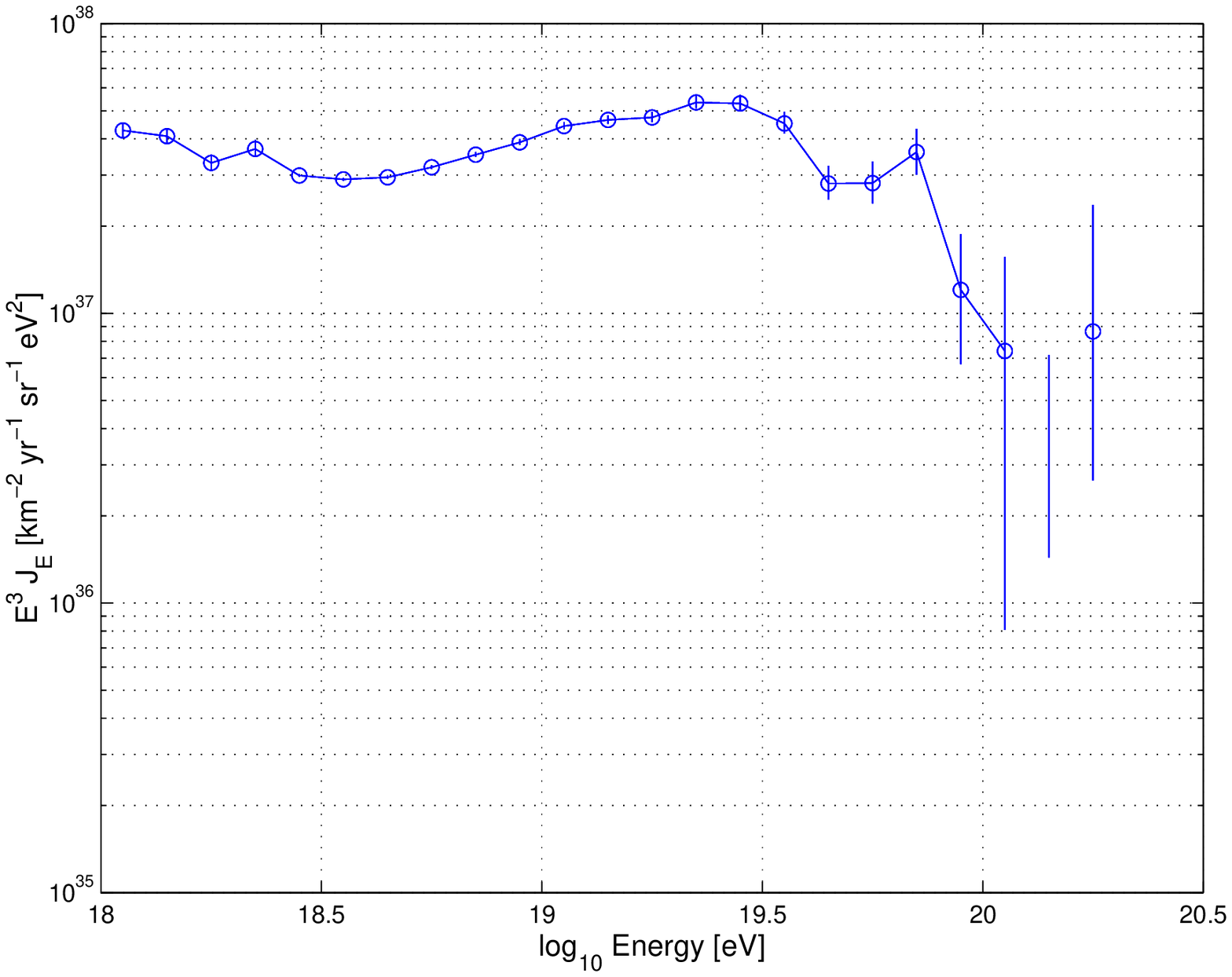}}
\caption{($Left.$) The energy spectrum of UHECRs in $10^{18-20}$ eV
by the Pierre Auger Observatory satisfies $J_E\propto E^{-3}$ to leading
order. ($Right.$) An ankle at about $10^{18.5}$ eV is apparent in $E^3J_E$, 
recently traced to the local mass distribution \cite{pao09}. The vertical bars 
refer to upper limits of statistical uncertainties, and the flux at $E=10^{20.15}$ eV 
is constrained only by an upper limit. (Courtesy Pierre Auger Collaboration.)}
\label{FIG_A1}    
\end{figure}

To leading order, the spectrum of cosmic rays is described by an intensity $J_E$ in 
eV$^{-1}$ km$^{-2}$ yr$^{-1}$ sr$^{-1}$, satisfying \cite{hor06}
\begin{eqnarray}
J_E\propto E^{-3}
\end{eqnarray}
over some 10 orders in energy $E$, further shown in Fig. \ref{FIG_A1}. 
An ankle exists at a few EeV (1 EeV=10$^{18}$ eV),
and knees at 4 and 400 PeV (1 PeV = $10^{15}$ eV) energies \cite{hor06}. High energy
protons ($E>5.7\times 10^{19}$ erg) produce pions in interaction with CMB photons 
$\gamma_{CMB}$ across an optical depth of 50 Mpc \cite{gre66,zat66}. 
In the energy range $10^{18-20}$ eV the PAO reveals an {\em excess} beyond the GZK threshold 
as shown in Fig. \ref{FIG_A1}. It points to two components below and above the break, where
the latter has been traced to the local mass distribution by correlations to nearby AGN in
the V\'eron-Cetty and V\'eron Catalogue \cite{ver06}.

For GRBs, most gamma-rays are sub-MeV \cite{mes06}, where the spectrum of the
prompt emissions (not including afterglows) displays power-law behavior modified 
by an exponentially suppressed tail. Their time-averaged spectra are well-described 
by the Band function \cite{ban93},
\begin{eqnarray}
N(E)=\left\{\begin{array}{ll}
((\alpha-\beta)E_0)^{(\alpha-\beta)}E^{\beta} e^{\beta-\alpha}
& [E<(\alpha-\beta)E_0]\\
E^\alpha e^{-E/E_0} & [E>(\alpha-\beta)E_0],
\end{array}
\right.
\end{eqnarray} 
where $(\alpha,\beta)$ denote low- and high-energy power-law indices
subject to $\alpha>\beta$ and $E_0$ within 100-1000 keV. Part of the scatter 
in $E_0$ is due to the broad redshift distribution of long GRBs, which peaks 
between $z=1$ and $z=2$, coincident with the maximum in the cosmic star formation rate. 
Corrected for redshift, the peak energies in GRBs cluster around 1 MeV \cite{loy02}. 
Likewise, $(\alpha,\beta)$ are broadly distributed with $\alpha<1$ 
(typically $\alpha<0$ with a mean of -1, \cite{pre00}) and $\beta<0$ 
\cite{ban93} (typically $\beta<-2$ \cite{pre00}). The Band spectrum is 
representative for the featureless optically thin synchrotron radiation
produced in collisionless shock fronts, possibly augmented with additional 
inverse (self-)Compton scattering \cite{mes06,wil09} and an thermal component 
\cite{per07}.

The recently launched {\em Fermi}/Glast satellite extends our window to
8 keV - 40 MeV by the Glast Burst Monitor (GBM) and 20 MeV - 300 GeV by the 
Large Area Telescope (LAT). It detected the most energetic event ever, 
GRB080916C \cite{abd09} of duration $T_{90}=66$ s with an isotropically equivalent 
energy output $E_{iso}=8.8\times 10^{54}$ erg at a redshift of 4.25 \cite{gre09} 
with one photon at 13 GeV (70 GeV in its restframe). Here, $T_{90}$ denotes the
90\% percentile of the integrated photon count of a GRB event. The Band parameters 
$(E_0,\alpha,\beta)$ show temporal behavior around canonical values 
$(500 \mbox{keV},-1,-2.2)$ and imply a minimum Lorentz factor 
$\Gamma_j=890$ \cite{taj09}.
\begin{figure}
\centerline{
\includegraphics[height=80mm,width=100mm]{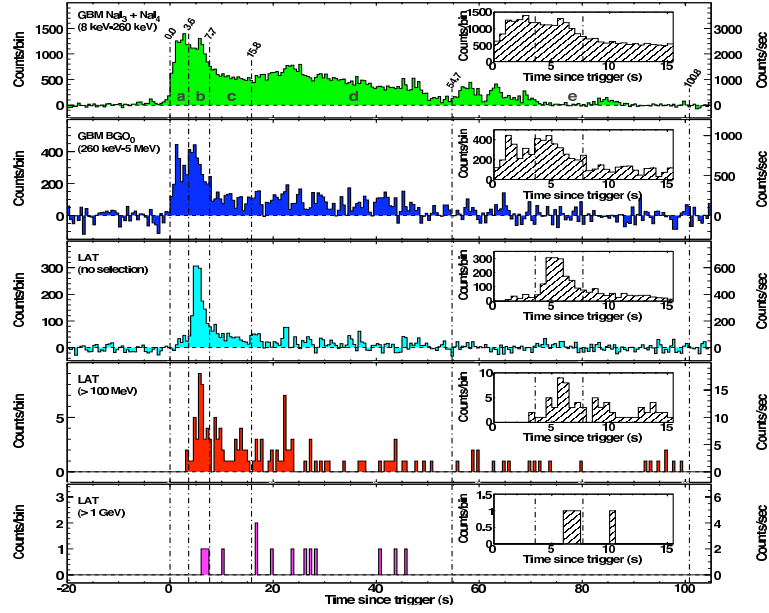}}
\caption{The light curves of GRB080916C ($T_{90}=66$ s, $z=4.25$) in four energy 
bands of {\em Fermi}/Glast: GBM 8-260 keV, GBM 0.26-5 MeV, LAT, LAT$>$100 MeV 
and LAT $>$ 1 GeV (top-to-bottom). The inserts show the distribution of photon 
counts in the first 15 s of the burst since the trigger. The results indicate a 
time delay of about 5 to 15 s in the highest energy photons received. (Reprinted 
with permission from \cite{abd09,taj09}. (c)2009 American Association of the Advancement 
of Science.)} 
\label{FIG_A2}    
\end{figure}

In anisotropic emissions, the true energy in gamma-rays satisfies $E_\gamma=f_b^{-1}E_{iso}$, 
where $f_b=2/\theta_j^2$ denotes the geometrical beaming factor for a two-sided 
jet-like outflow with half-opening angle $\theta_j$, and $E_{iso}$ denotes the observed
isotropic equivalent emissions (ignoring anisotropy). Many GRB light curves 
display an achromatic break in their light curves, when the relativistic 
beaming angle $\sim 1/\Gamma_j$ of radiation produced by shock fronts in the
jet exceeds the half-opening angle $\theta_j$ of the jet \cite{mes06}. No 
clear measurement of $\theta_j$ has been made for GRB080916C \cite{gre09}. 
Nevertheless, we may estimate $E_\gamma\sim 10^{52}$ erg based on the average 
$f_b\sim 500$ \cite{fra01,van03b}. GRB080916C reveals a relatively late 
onset of maximal flux in the highest energy photons, shown in Fig. \ref{FIG_A2}.

Direct evidence for magnetic fields in GRB emissions remains difficult in the 
absence of unambiguous polarization measurements. Identification of the constituents 
also remains an open challenge. (The latter applies to extragalactic radio jets 
as well.) {\em Fermi}-LAT conceivably offers a window to synchrotron emissions 
from protons, rather than electron-positrons alone \cite{asa09,asa09b}.

\begin{figure}
\centerline{
\includegraphics[angle=270,scale=0.30]{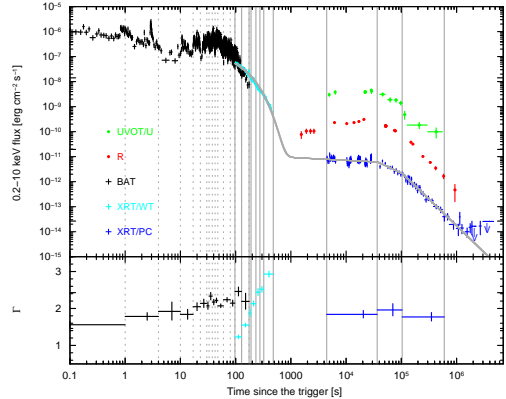}}
\caption{{\em Swift} observation of GRB060614 ($T_{90}=102$ s, $z=0.125$, 
$E_\gamma\simeq 4.2\times 10^{49}$ erg) and its extended tail in X-ray emissions 
($t>$ 1 ks). The latter features a plateau $(L_X\simeq 3\times 10^{41}$ erg s$^{-1}$) 
followed by exponential decay with e-folding timescale $\tau=76$ s. The lower panel shows 
the photon index in the Burst Alert Telescope (BAT, 35-350 keV) and X-Ray Telescope (XRT, 
0.2-10 keV) energy bands. (Reprinted with permission from \cite{man07}. (c)2007 ESO)}
\label{FIG_A3}    
\end{figure}

Recently, {\em Swift} identified extended X-ray tails to the prompt GRB emissions
on a timescale of 1 - 1000 ks \cite{zha07a,zha07b}, shown in Fig. \ref{FIG_A3} for 
the unique event GRB060614 with no supernova \cite{del06,geh06}. XRTs are common to long 
($T_{90}>2$ s) and short GRBs ($T_{90}<2$ s) events \cite{man07,eic09} with apparently 
no memory of the initial state giving rise to the prompt GRB emissions. For GRB060614,
they are described by broken power laws with a photon index of about 1.6 \cite{man07}.

Summarizing, UHECR and GRB observations pose a number of challenges which may reveal
processes, whereby non-thermal radiation is produced by strong gravity around black holes. 
Because of the universality of gravitational interactions, we hereby anticipate a broad 
range of radiative signatures in particles (hadrons and leptons) and photons alike, and 
possibly so from a single source. Such multi-messenger sources are a focal point of
{\em astroparticle physics}, led by the emerging multi-spectrum surveys of the sky. In 
particular, we mention the challenges to understand the
\begin{enumerate}
\item
Physical mechanism for accelerating UHECRs up to and beyond the GZK threshold and the
energy reservoir powering this process and their hadronic composition, tentatively
associated with low-luminosity Seyfert galaxies;
\item 
Universality of GRB inner engines of short and long durations across a diversity in 
astrophysical progenitors and host environments, with and without supernovae 
(GRB060614) and/or pronounced X-ray afterglows (GRB050911 \cite{pag06}), ocurring
in and away from star forming regions, respectively \cite{pac98};
\item
Physical mechanism for producing ultra-relativistic leptonic and/or electromagnetic
outflows (baryon-poor jets) powering long and short GRBs and the energy reservoir powering
this output. For inner engines harboring stellar mass black holes, the observed true 
luminosity in gamma-rays exceeds the Eddington luminosity by a factor of about $10^{10}$ 
on the basis of \cite{fra01};
\item 
Mechanism whereby long lived inner engines produce aspherical and radio-loud supernovae 
in core-collapse events producing long GRBs \cite{hoe99}; and the observational 
consequences of the same energy output in long GRBs sans supernova;
\item 
Remnant of the GRB inner engine, powering prolonged X-ray tails common to long and short GRBs;
\end{enumerate}
and, in a latest twist to GRB phenomenology, the physical mechanism for a delayed peak luminosity 
in the highest energy photons (GRB080916C).

\section{Physical properties of Kerr black holes}

The Kerr metric \cite{ker63} is an exact two-parameter solution of the metric in general 
relativity as a function of mass, $M$, angular momentum, $J$. The Kerr metric is preserved 
when it accumulates a small electric charge, $Q^2/M<<1$. Its energy reservoir is like
that of a spinning top,
\begin{eqnarray}
E_{rot} = \frac{1}{2}I\Omega_H^2,
\label{EQN_rot1}
\end{eqnarray}
where $I\simeq 4M^3$ \cite{tho86} denotes the moment of inertia in the limit of small
angular velocities $\Omega_H$.

Angular momentum gives a non-zero contribution to the Riemann tensor, as described by the
Kerr metric. It comes with a converse: a local force on a test particle arising from the 
product of angular momentum $J$ [cm$^2$] and the Riemann tensor [cm$^{-2}$], first described 
by Papapetrou \cite{pap51,van05,van08a,luc08} and here expressed in geometrical units. The 
result is an exchange of force between objects with angular momenta $J_{1,2}$ via
\begin{eqnarray}
J_1\leftrightarrow{\mbox{Riemann}}{\leftrightarrow} J_2,
\label{EQN_J12}
\end{eqnarray}

Electromagnetic radiation is created by charged particles. In the magnetosphere around black 
holes \cite{bla77,tho86}, charged particles carry an appreciable canonical angular momentum
and, by (\ref{EQN_J12}), should give rise to non-thermal radiation processes \cite{van09}.

\subsection{Spin energy}

We can generalize (\ref{EQN_rot1}) to arbitrary spin rates of a Kerr black hole
using $\sin\lambda=a/M$ \cite{van99}, where $a$ denotes the specific angular momentum 
$J/M$, as
\begin{eqnarray}
E_{spin}=\frac{1}{2}I\Omega_H^2f_s^2 = 2M\sin^2(\lambda/4),
\label{EQN_rot2}
\end{eqnarray}
where $f_s=\frac{\cos(\lambda/2)}{\cos(\lambda/4)}$ and $\Omega_H=\frac{1}{2M}\tan(\lambda/2)$
\cite{van99}. The ratio
\begin{eqnarray}
\frac{E_{spin}}{\Omega_HJ}=\frac{1}{2}\cos^{-2}(\lambda/4)
\label{EQN_TOP}
\end{eqnarray}
satisfies
\begin{eqnarray}
\frac{1}{2} \le \frac{E_{spin}}{\Omega_HJ} \le 0.5858 ~~~\left(0\le \lambda \le\frac{\pi}{2}\right).
\end{eqnarray}
It remains remarkably close to the Newtonian value 1/2, highlighting
a close similarity of Kerr black holes with spinning tops. Note that (\ref{EQN_rot2}) 
involves {\em no} small parameter, whereby $E_{spin}/M$ can reach 29\%. This is an 
order of magnitude larger than the spin energy of a neutron star. Table \ref{TAB1} 
gives a list of the physical properties of Kerr black holes.

\begin{table}
\center{
\begin{tabular}{lclcr}
{\sc Symbol}  &~~~~~~~~~~~~~~~~~~~~~& {\sc Expression}  &~~~~~~~~~~~~~~~~~~~~~& {\sc Comment} \\
\hline
$\lambda$     && $\sin\lambda=a/M$               && rapidity of spin\\
$r_H$ 		  && $2M \cos^2(\lambda/2)$          && horizon radius\\
$\Omega_H$	  && $\frac{1}{2M}\tan(\lambda/2)$ 	 && angular velocity\\ 
$E_{spin}$ 	  && $2M\sin^2(\lambda/4)$           && spin-energy, $\le0.29M$\\
$M_{irr}$	  && $M\cos(\lambda/2)$ 	         && irreducible mass, $\ge 0.71M$\\
$A_H$ 		  && $16\pi M^2_{irr}$	             && surface area\\
$S_H$		  && $\frac{1}{4}A_H$                && entropy\\
$T_H$ 		  && $\frac{1}{4r_H}\cos\lambda$     && temperature\\
$\mu_H$		  && $QJ/M$		                     && Carter's magnetic moment\cite{car68}\\
$Q_e$		  && $\simeq 2BJ$		             && Wald's equilibrium charge\cite{wal74}\\
$V_F(\theta)$ && $e\Omega_HA_\phi(\theta)$       && Fermi-level at poloidal angle $\theta$ \cite{van00}\\ 
\hline
\end{tabular}
}
\caption{Physical properties of Kerr black holes of mass $M$ and angular momentum 
$J=M^2\sin\lambda$ are listed in geometrical units (Newton's constant and the velocity 
of light $c=1$) and $\hbar=1$. Exposed to an axisymmetric vector potential $A_a$ with 
magnetic field-strenth $B$, there exists an equilibrium charge $Q_e$ and a horizon 
Fermi-level $V_F$, expressed for particles of charge $e$.}
\label{TAB1}
\end{table}

In astrophysical environments, the evolution of Kerr black holes satisfies the 
first law of thermodynamics \cite{bar73},
\begin{eqnarray}
dM = \Omega_H dJ + T_H dS_H,
\label{EQN_1st}
\end{eqnarray}
where $T_H$ denotes the temperature of the event horizon, associated with a change
$\Omega_H dJ$ in spin energy and a dissipation $T_H dS_H$ with creation of 
Bekenstein-Hawking entropy \cite{bek73}
\begin{eqnarray}
S_H = 4\pi M^2\cos^2(\lambda/2).
\end{eqnarray}
Thus, $S_H$ can maximally {\em double} in the process of viscous spin-down 
of an initially rapidly spinning black hole to a Schwarzschild black hole.

Rotating black holes develop a lowest energy state, described by an equilibrium charge 
$Q=2BJ$ in an asymptotically uniform magnetic field of strength $B$ aligned with its spin 
axis \cite{wal74}. It preserves essentially maximal magnetic flux through the event horizon 
at arbitrary rotation rates, and can be seen, within a factor of two, to arise as the minimum 
$Q=BJ(r_H/M)$ of the potential energy \cite{van01} ${\cal E} \simeq \frac{1}{2} C Q^2 - \mu_H B,$
where $C\simeq 1/r_H$ denotes the electrostatic capacitance for of a black hole 
of size $r_H$ and $\mu_H=QJ/M$ by Carter's theorem \cite{car68}. The Carter's magnetic moment
in equilibrium hereby satisfies
\begin{eqnarray}
\mu_H^e\simeq 2\frac{BJ^2}{M}.
\label{EQN_MUH}
\end{eqnarray}

\subsection{Frame dragging}

Mach observed a coincidence of zero angular velocity with respect to 
infinity and zero angular momentum of an object. He attributed it 
to dragging of space with matter, recognizing that most mass is in the 
distant stars. It defines a {\em superposition principle}, whereby dragging 
is a net result summed over contributions from matter distributed over space. 
Applied to spaces with black holes, dragging will be led by the mass-energy 
of a black hole nearby and by the distant stars far out. The result is described 
by {\em frame dragging}, which reaches the angular velocity of the black hole on 
its event horizon and which vanishes at large distances. It is differential, and 
not a gauge effect.

The Kerr metric gives an exact solution of frame dragging in terms of the
angular velocity $\omega$ of particles of zero-angular momentum. It is 
parametrized by the mass and angular momentum of the black hole.
In Boyer-Lindquist coordinates $(t,r,\theta,\phi)$ of the 
Kerr metric \cite{tho86}, the world line of zero-angular momentum particles are 
orthogonal to slices of constant time-at-infinity, and their angular velocity 
$\omega=d\phi/dt$ decays with the cube of the distance to the black hole at large 
distances.
\begin{figure}
\centerline{
\includegraphics[scale=.25]{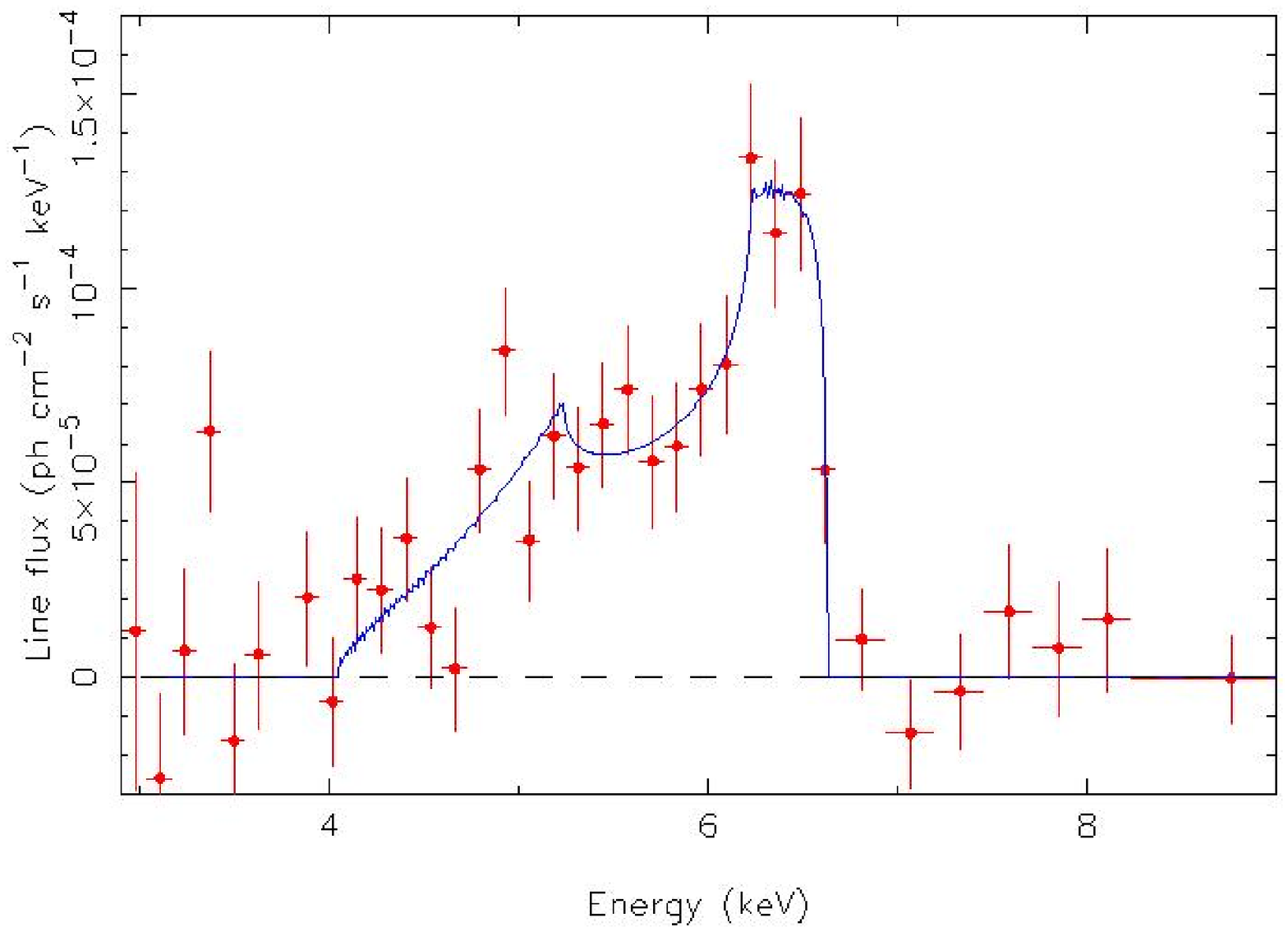}\includegraphics[scale=.26]{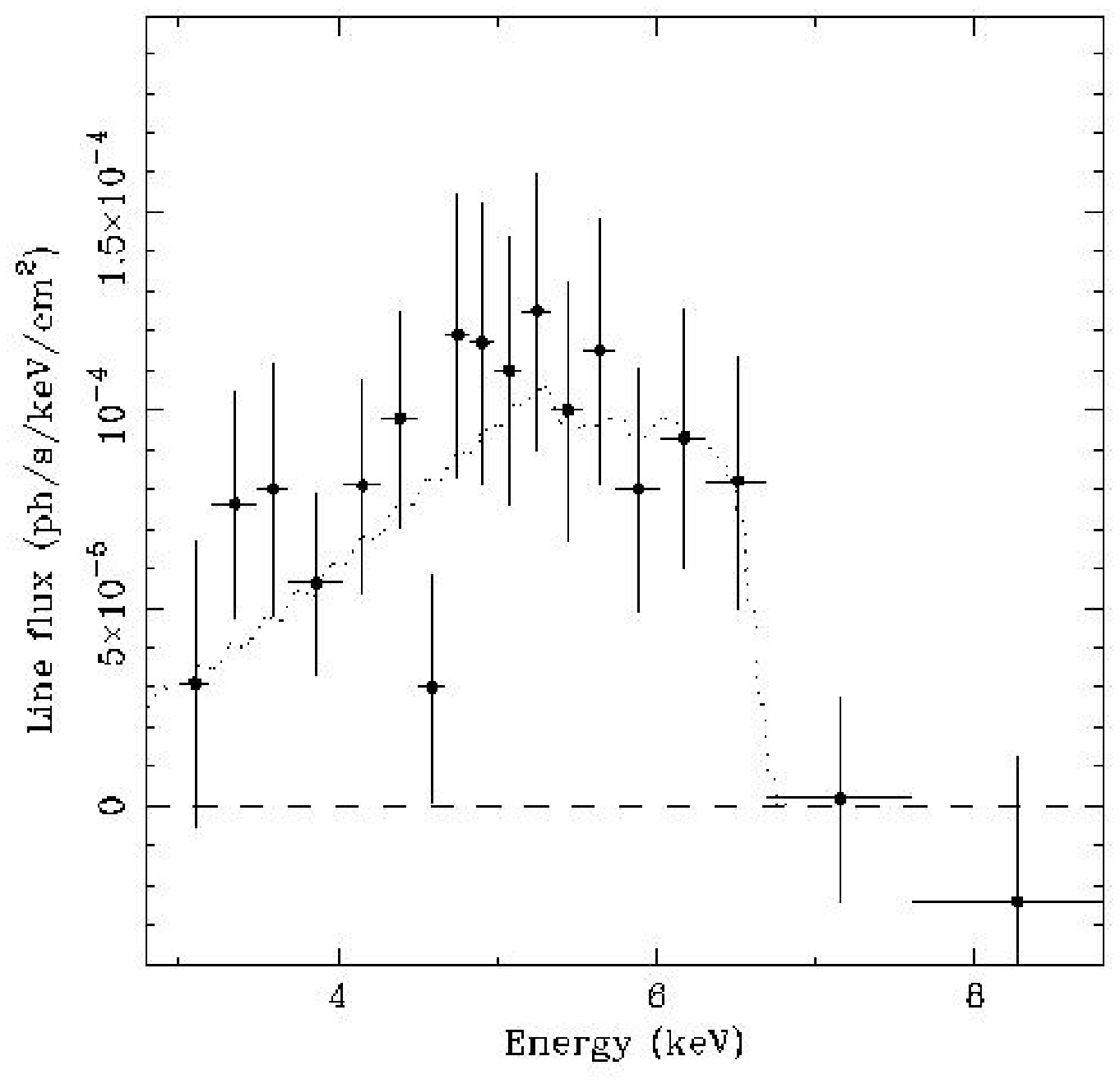}}
\caption{The K$\alpha$ iron line-emissions in the Seyfert galaxy MCG 6-30-15 
observed by ASCA reveal relativistic orbital velocities in asymmetrically 
redshifted and blueshifted peaks around the rest frame energy of 6.35 keV. It 
is consistent with line-emissions from a disk subject to redshift and Doppler 
shifts in orbital motion around a supermassive black hole, where the cut-off at 
4 keV corresponds to truncation of a disk at radius $6M$. (Reprinted with permission 
from \cite{tan95}. (c)1995 McMillan Publishers.) In a different epoch, it displayed 
an extended tail well below 4 keV associated with matter in stable orbital motion 
inside $6M$, indicative of a compact ISCO around a rapidly rotating black hole, with
a noticeably lower luminosity in blueshifted emissions. (Reprinted with permission 
form \cite{iwa96}. (c)1996 The Royal Astronomical Society.)}
\label{FIG_MCG}    
\end{figure}

The angular velocity of matter in circular orbits around Kerr black holes 
satisfies \cite{sha83}
\begin{eqnarray}
\Omega_T=\pm \frac{1}{z^{3/2}\pm \hat{a}}
\label{EQN_OMT}
\end{eqnarray}
for co-rotating (+) and counter-rotating (-) orbits, where $z=r/M$ and $\hat{a}=a/M$.
frame dragging changes the radius of the inner most stable circular orbit (ISCO), where
the specific energy and angular momentum of test particles satisfy \cite{sha83}
\begin{eqnarray}
e=\sqrt{1-\frac{2}{3z}},~~l=\frac{2M}{3\sqrt{3}}\left(1+2\sqrt{3z-2}\right),
\label{EQN_EL}
\end{eqnarray}
where $z= 3 + Z_2 \mp \left[(3-Z_1)(3+(Z_1+2Z_2)\right]^{1/2}$
in terms of $Z_1=1+(1-\hat{a}^2)^{1/3}$ $\left[(1+\hat{a})^{2/3}+(1-\hat{a})^{1/2}\right]$,
$Z_2=(3\hat{a}^2+Z_1^2)^{1/2}$. We note that $l/e$ decreases from $3\sqrt{3/2}M$ for
$\hat{a}=0$ $(z=6)$ down to $l/e=2M$ for $\hat{a}=1$ ($z=1$). In dimensionful form,
the specific angular momentum $j$ of particles in stable circular orbits satisfies
\begin{eqnarray}
j\ge lGM/c,
\label{EQN_j}
\end{eqnarray}
where $2/\sqrt{3}<l<2\sqrt{3}$ \cite{sha83}.
Thus, frame dragging allows the ISCO to shrink from $6M$ around a Schwarzschild black hole 
to $M$ around an extremal Kerr black hole for corotating orbits. Its effect on the ISCO 
can be measured with X-ray spectroscopy, as shown in Fig. \ref{FIG_MCG}. The result for
MCG 6-30-15 is notably time variable, revealing distinct spectra taken at different 
epochs \cite{iwa96}. 
It may reflect intermittency in the inner radius of the disk \cite{fab95} or,
alternatively, in circumnuclear clouds intermittently absorbing disk emissions.

The complete gravitational field induced by the angular momentum and mass of a rotating black hole
is encoded in the Riemann tensor of the Kerr metric, enabling some exact results on (\ref{EQN_J12}).

\section{Black hole radiation processes: some general considerations}

In modeling black hole radiation proceses, we are led to consider (1) causality,
(2) the loading problem, (3) spectral energy correlations and light curves 
associated with the state of accretion and their Eddington luminosities, as
well as observational tests against existing and future observations. 

In black hole spin driven radiation processes, causality is non-trivial in view of the 
ingoing radiation boundary conditions on the event horizon \cite{tho86}. Causality can
be satisfied by gravitational interactions onto surrounding fields and matter, for
inducing radiative processes in which the event horizon assumes a passive role. Here,
the interactions may appear in integral form, but are preferrably expressed by a 
principle of equivalence or in local interactions arising from spacetime curvature.

The loading problem refers to the identification of observable emissions and their fractional 
output in luminosity or total energy output (``the energy budget" of an inner engine). In 
light of observations on UHECRs and GRBs, these emissions are notably non-thermal, generally at 
high energies in open outflows and low energies from matter surrounding the black hole. It 
poses questions on the physical mechanisms for converting spin energy into, e.g., leptonic 
and hadronic output as well as various emission channels via surrounding matter.

GRBs represent single, non-repeating burst sources. For these sources, we can derive 
spectral energy correlations \cite{ama02,ghi04,ghi07,van08a} covering the entire burst  
and extract an average light curve, following normalization of durations and count rates 
\cite{van09}. The results, based on data from HETE II, Swift and BATSE, can be used for 
detailed confrontation with spectra and light curves from models. Model light curves and the 
Eddington luminosity of high energy emissions inevitably depend on the state of accretion, 
whereby a detailed comparison with data of the former promises to define a novel probe of 
the latter. Similar but not identical considerations apply to AGN in view of their extended
lifetimes.

An early Ansatz for modeling energy extraction of black hole spin of black holes in
astrophysical environments is based on horizon Maxwell stress \cite{ruf75} and ample
electron-positrons in the black hole environment for it to assume a force-free state \cite{bla77}. 
Analysis of a global force-free solution around slowly spinning black holes with
vanishing Carter's magnetic moment (\ref{EQN_MUH}) points to a 
net energy output in a state of accretion \cite{bla77}.

However, global, steady-state force-free solutions do not elucidate causality \cite{pun90},
while accretion leads to open models wherein essentially all black hole output is
channeled into an open Poynting flux \cite{van01}. The limit in which (\ref{EQN_MUH}) is
zero limits applicability to slowly spinning black holes, wherein accretion serves to supply 
a constant magnetic flux onto the event horizon. Accretion tends to produce continuous spin 
up of the black hole \cite{kum08}, upon neglecting any backreaction of black hole spin onto 
the surrounding matter and its contemporaneous emissions. The lifetime of black 
hole spin is hereby indefinite.

In what follows, we take, instead, frame dragging as a universal starting point for 
first-principle interactions. It serves 
as an ab initio causal agent for radiation processes around black holes in equilibrium 
(\ref{EQN_MUH}) with arbitrary spin rates. We aim for a local description for
frame dragging induced interactions with particles and the environments, subject to a 
direct confrontation with multi-messenger data on durations, energies and light curves,
present and upcoming.

\section{Frame dragging induced radiation processes}

Frame dragging is a powerful agent in the absence of a small parameter in the proximity
of the black hole for inducing energetic interactions.
Its manifestation is along the spin axis of the black hole by (\ref{EQN_J12}) 
and in the interaction with matter via a torus magnetosphere.

\subsection{Gravitational spin-orbit energy $E=\omega J_p$}

From a position $r$ in Boyer-Lindquist coordinates along the spin-axis of a black hole, 
the line-integral of the Papapetrou force in the Kerr metric defines a
potential energy \cite{van05,van08a}
\begin{eqnarray}
 {\cal E}(r,\theta)=\int_r^\infty \mbox{Riemann}\times J_p ds = \omega(r,\theta) J_p,
\label{EQN_E1}
\end{eqnarray}
where $\omega(r,\theta)$ refers to the frame dragging angular velocity at $(r,\theta)$, 
and $J_p$ denotes the conserved angular momentum of the test particle. Applied to a 
charged particle, $J_p=eA_\phi$ along an open tube with magnetic flux $2\pi A_\phi$, 
${\cal E}$ can be arbitrarily large, depending on the magnetic field-strength and the size 
of the black hole.
\begin{figure}
\centerline{
\includegraphics[scale=.40]{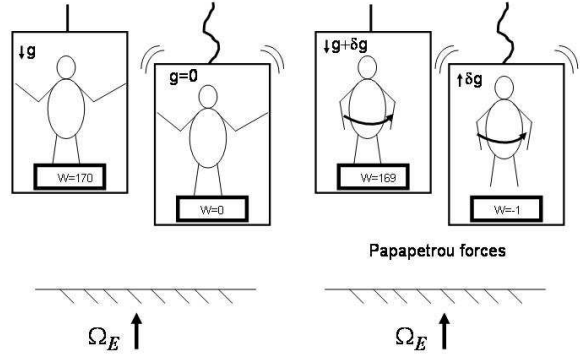}}
\caption{The (weak) equivalence principle gives rise to weightlessness in a freely falling 
elevator, apparent in zero relative acceleration $(g=0$). Higher-order interactions ($\delta g\ne0$) 
exist between the angular momentum of the passenger and the angular momentum of the Earth via the  
Riemann tensor as described by the Kerr metric, giving rise to a Papapetrou force. Corotation 
between the passenger, like a ballerina, and the Earth, with angular velocity $\Omega_E$, hereby 
reduces weight, leading to an apparent lift-off $(\delta g>0$) of the passenger in a freely falling 
elevator. The opposite result is obtained for counter-rotation between passenger and the Earth.}
\label{FIG_MX}    
\end{figure}

We shall derive (\ref{EQN_E1}) in the approximation of small $\theta$ in the
Kerr metric. In a frame of tetrad 1-forms
\begin{eqnarray}
e_{(0)}= \alpha dt,~~ 
e_{(1)} =\frac{\Sigma}{\rho}(d\phi - \omega dt)\sin\theta,~~
e_{(2)}= \frac{\rho}{\sqrt{\Delta}} dr,~~
e_{(3)} = \rho d\theta,
\end{eqnarray}
where $\alpha={\rho}{\Sigma}^{-1}\sqrt{\Delta}$ denotes the redshift factor, 
$\Sigma^2=(r^2+a^2)^2-a^2\Delta\sin\theta$, $\rho=r^2+a^2\cos^2\theta$,
$\Delta=r^2-2Mr+a^2$ and $\omega={2aMr}{\Sigma}^{-2}$ denotes the angular 
velocity of frame dragging, the non-zero components of the Riemann tensor are \cite{cha83}
\begin{eqnarray}
\begin{array}{rcl}
R_{0123} &=& A,~~R_{1230} = AC,~~R_{1302} = AD\\
-R_{3002} &=& R_{1213} = -3aA\sqrt{\Delta}\Sigma^{-2}(r^2 + a^2)\sin\theta\\
-R_{1220} &=& R_{1330} = -3aB\sqrt{\Delta}\Sigma^{-2}(r^2 + a^2)\sin\theta\\
-R_{1010} &=& R_{2323} =  B = R_{0202} + R_{0303}\\
-R_{1313} &=& R_{0202} =  BD,~~-R_{1212} = R_{0303} = -BC,
\end{array}
\label{EQN_R}
\end{eqnarray}
where
\begin{eqnarray}
\begin{array}{rcl}
A &=& aM\rho^{-6}(3r^2 - a^2 \cos^2\theta),~~B = Mr\rho^{-6}(r^2 - 3a^2 \cos^2\theta),\\
C &=& \Sigma^{-2}[(r^2 + a^2)^2 + 2a^2\Delta\sin^2\theta],~~D = \Sigma^{-2}[2(r^2 + a^2)2 + a^2\Delta\sin^2\theta].
\end{array}
\end{eqnarray}
Along the spin-axis axis of the black hole ($\theta=0$),
$2A = -\partial_r\omega = {2aM}{\rho^{-6}}(3r^2 - a^2)$, $C = 1$, $D = 2,$
giving rise to black hole-spin induced curvature components in the 
first three of (\ref{EQN_R}). The Papapetrou force \cite{pap51}
\begin{eqnarray}
F_2 =\frac{1}{2}\epsilon_{abef}R^{cf}_{cd}J_p^au^bu^d = J_pR_{3120} = J_pAD = -\partial_2\omega J_p,
\end{eqnarray}
where $u^b$ denotes the velocity four-vector of the test particle, can be integrated out to 
infinity,
\begin{eqnarray}
{\cal E}=\int_r^\infty F_2 ds,
\label{EQN_E2}
\end{eqnarray}
giving (\ref{EQN_E1}).

The equation (\ref{EQN_E2}) can also be derived as follows. Let $u^b$ denote the velocity 
four-vector and $\Omega=u^{\phi}/u^t$ denote their angular velocities. Normalization
$-1 = u^cu_c = \left[g_{tt} + g_{\phi\phi}\Omega(\Omega-2\omega)\right](u^t)^2$
gives two roots $\Omega_{\pm} = \omega\pm\sqrt{\omega^2 - (g_{tt} + (u^t)^{-2})/g_{\phi\phi}}.$
For two particles with the same angular momentum in strenght,
\begin{eqnarray}
J_{p,\pm} = g_{\phi\phi}u^t(\Omega_{\pm}+\omega) = 
  g_{\phi\phi}u^t\sqrt{\omega^2-(g_{tt}+(u^t)^{-2})/g_{\phi\phi}}=\pm J_p
\end{eqnarray}
we find the same $u^t$ for each particle. Their total energy satisfies 
$E_{\pm} = (u^t)^{-1} + \Omega_{\pm}J_{\pm},$ and hence one-half the difference satisfies
\begin{eqnarray}
{\cal E} =\frac{1}{2}(E_+ - E_-) = \omega J_p.
\label{EQN_E3}
\end{eqnarray}

The curvature-spin coupling (\ref{EQN_E1}) is gravitational, i.e., it is universal 
irrespective of whether the angular momentum is mechanical or electromagnetic in origin.

\subsection{Gravitational interaction ${\cal L}_\omega B$ with a torus magnetosphere}

\begin{figure}
\centerline{
\includegraphics[scale=.34]{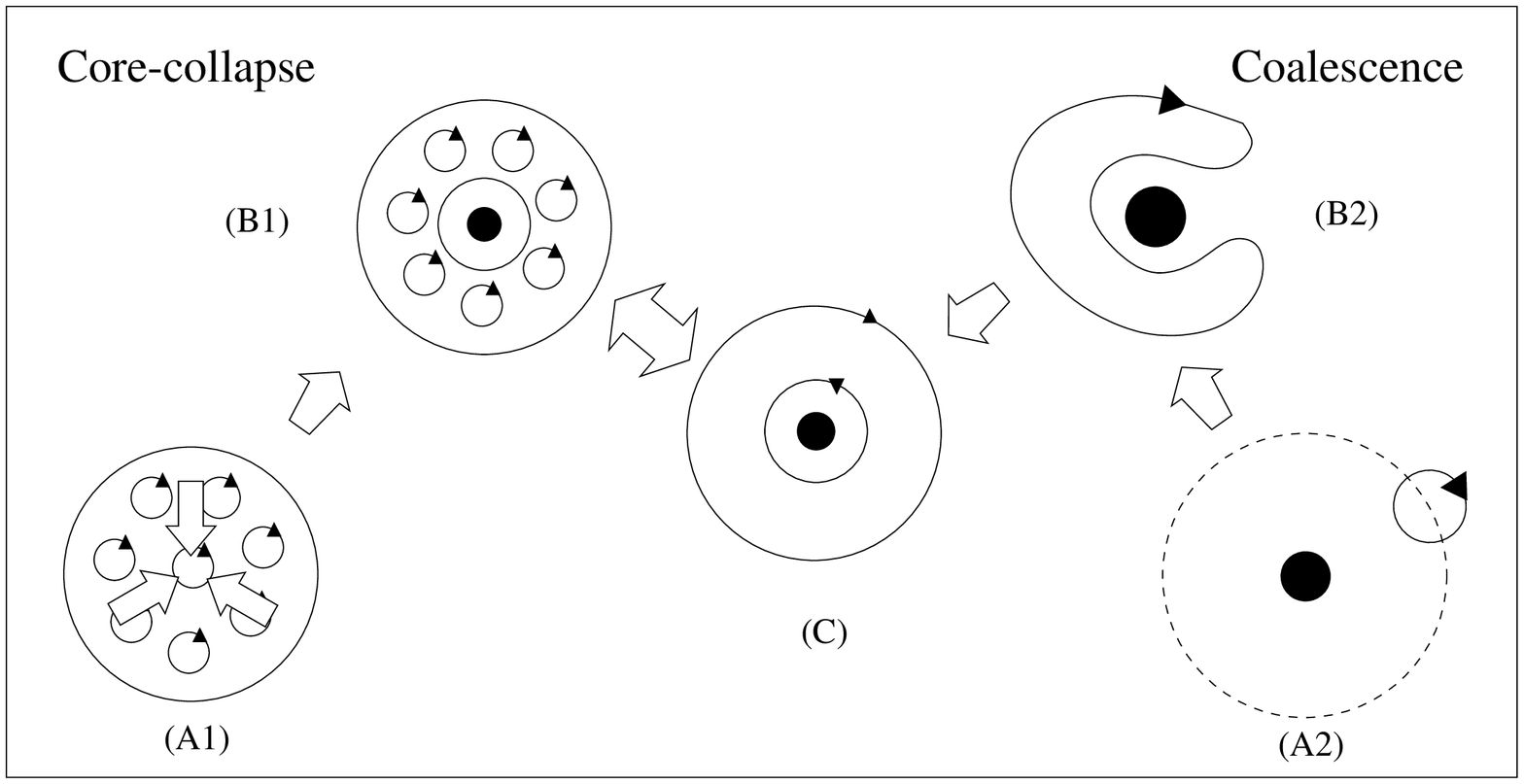}\includegraphics[scale=.34]{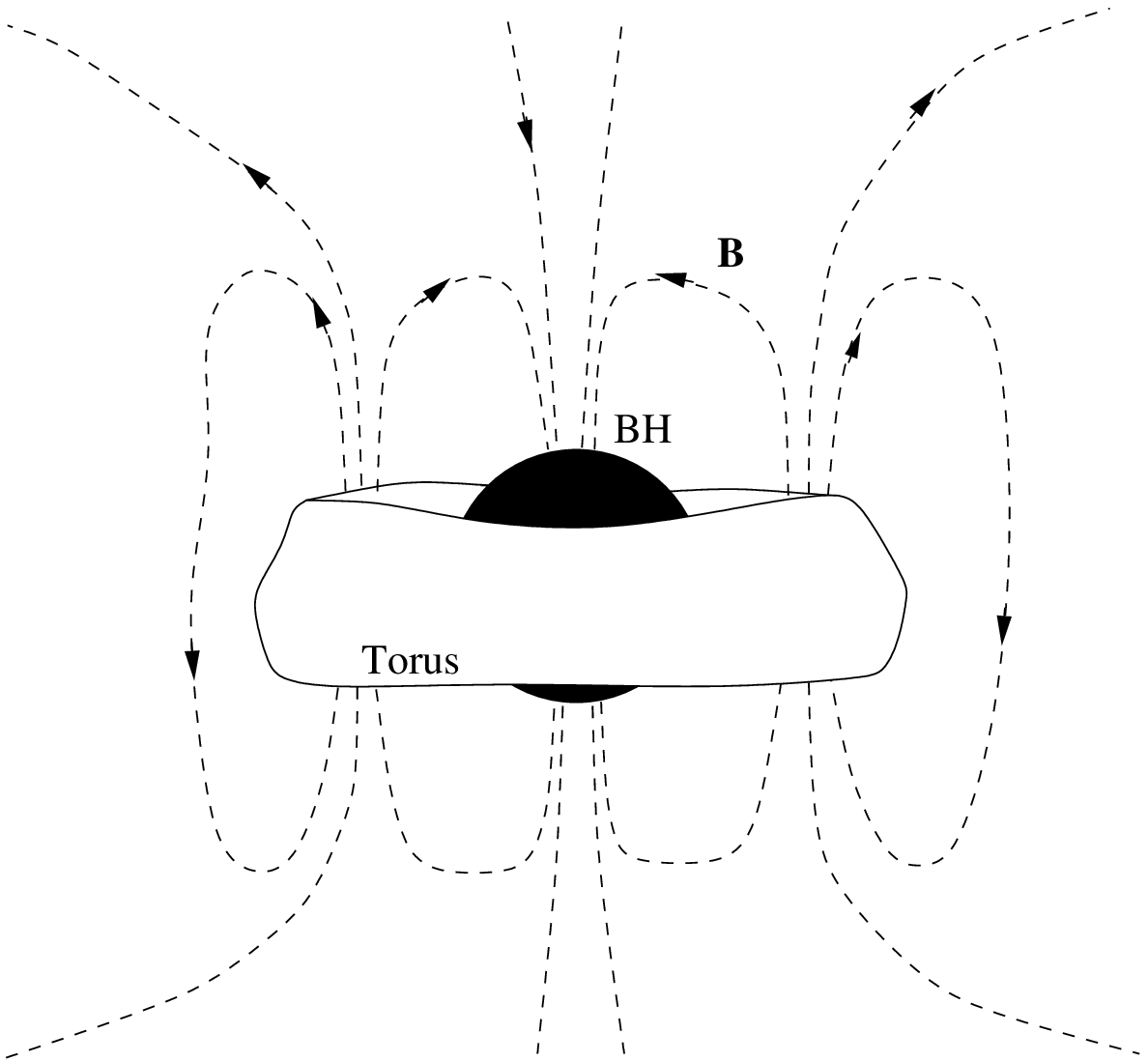}}
\caption{($Left.$) Kerr black holes surrounded by a torus are expected to be formed in 
core-collapse of massive stars \cite{van04,woo93}, mergers of neutron stars with a 
stellar mass black hole \cite{pac91,van99}, or the merger of two neutron stars (\cite{bai08}, 
not shown). The topology of formation of the first two scenarios is shown in 
equatorial cross-sections (A1-C) and (B1-C). The instantaneous poloidal flux of an 
otherwise turbulent magnetic field is here represented by equatorial current loops or, 
equivalently by Stokes' Theorem, two counter-oriented current loops. 
(Reprinted from \cite{van03}. (c)2009 The American Astronomical Society.) ($Right.$) 
The outcome is a black hole surrounded by a torus magnetosphere which, in its lowest 
energy state, assumes an equilibrium magnetic moment (\ref{EQN_MUH}). It preserves
essentially maximal horizon flux at arbitrary spin rates and supports open magnetic 
flux lines to infinity. The result is a long-lived inner engine for the lifetime of
black hole spin during spin-down against the surrounding torus \cite{van99}. 
(Reprinted from \cite{van01}. (c)2001 The American Physical Society.)} 
\label{FIG_MX0}
\end{figure}

Accretion disks are believed to carry turbulent magnetic fields by the magnetorotational 
instability (MRI, \cite{bal91}). Of particular interest is the {\em infrared spectrum} of 
magnetohydrodynamical (MHD) turbulence at low azimuthal quantum number, $m$. The variance 
in the $m=0$ component of the poloidal magnetic field represents the energy in net poloidal 
flux, which establishes a torus magnetosphere surrounding the black hole also without 
accretion \cite{van99}. Some generic astrophysical realizations are illustrated in Fig. 
\ref{FIG_MX0}.

\begin{figure}
\centerline{
\includegraphics[scale=.30]{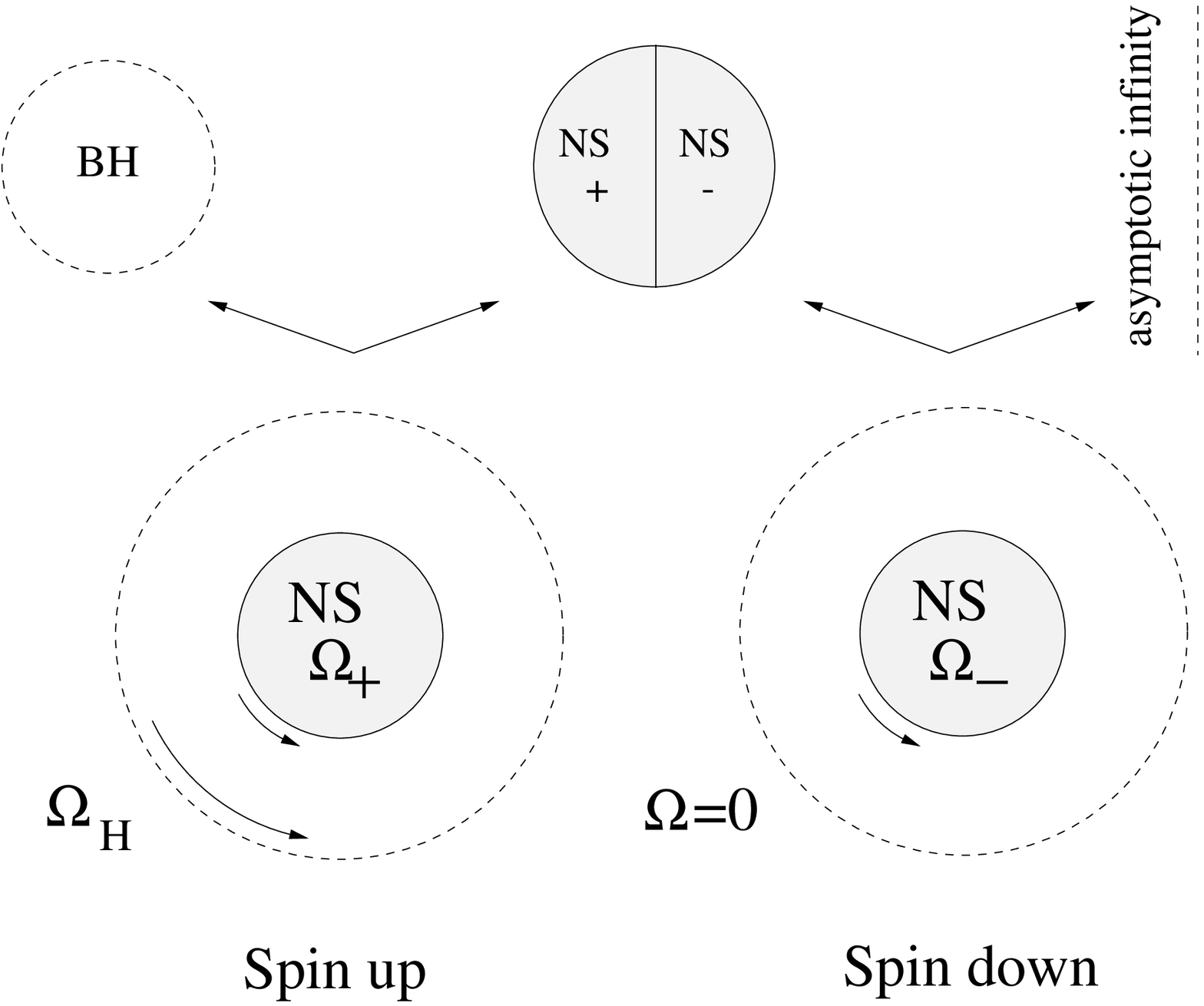}}
\centerline{\includegraphics[scale=.15]{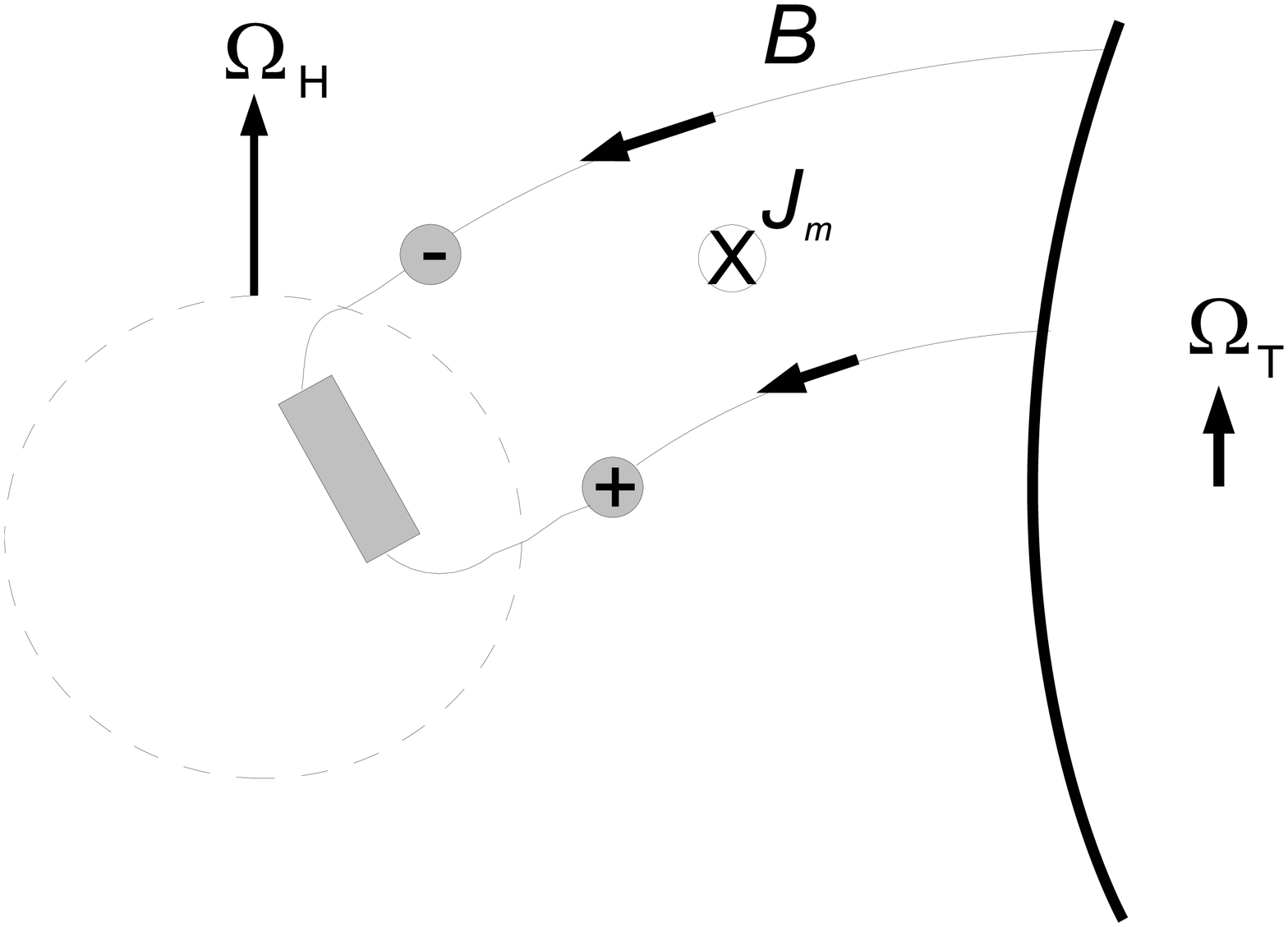}\includegraphics[scale=.15]{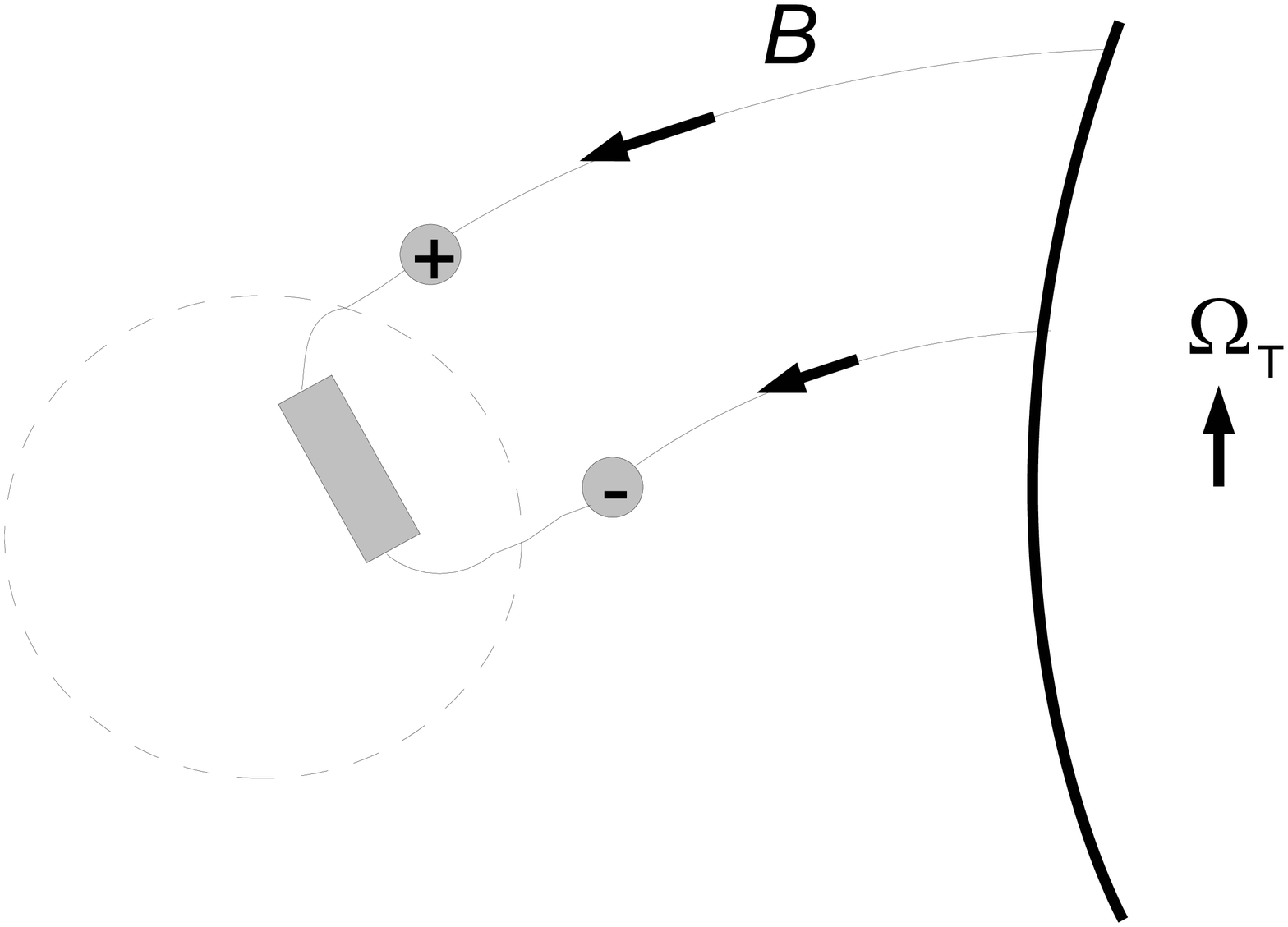}}
\caption{Mach's principle applied symmetrically to matter at asymptotic infinity and to
a rapidly rotating black hole nearby implies radiation of energy and angular momentum from the black 
hole via a surrounding torus. ($Top.$) In a poloidal cross-section, the two faces of the torus
can each be identified with neutron star (NS+,NS-) with angular velocities $\Omega_+,\Omega_-)$.
($Middle.$) The inner face (NS+) is subject to spin up, as does a (slowly rotating) neutron star when infinity 
wraps around it with angular velocity $\Omega_H$, shown in equatorial cross-section. 
The outer face (NS-) spins down like any neutron 
star by shedding energy and angular momentum in magnetic winds to infinity. 
({Reprinted from} \cite{van03}. (c)2003 The American Astronomical Society.) ($Bottom.$) 
The same result can be inferred from Faraday's equations in curved spacetime, applied to the 
inner and outer torus magnetosphere, shown in poloidal cross-section, subject to no-slip boundary 
conditions on the torus and radiative (slip) boundary conditions on the event horizon of the black hole and infinity. 
A frame dragging induced polarization shown on the terminals of the impedance elements arises out
of a current ${\cal J}_m$ (\ref{EQN_JM}),
as if produced by a virtual azimuthal current of 
magnetic monopoles. The black hole horizon hereby serves as a passive load. The induced 
poloidal currents mediate energy and angular momentum transfer by Maxwell stresses on the 
event horizon \cite{ruf75,bla77} and the inner face of the torus \cite{van99}.}
\label{FIG_MX1}    
\end{figure}

The angular velocity $\Omega_H$ of the event horizon of a rotating black hole can readily
exceed that of surrounding matter. Via an inner torus magnetosphere, matter can then receive input 
from the black hole in a process which is equivalent to the spin down of neutron stars by shedding 
energy and angular momentum to infinity (Fig. \ref{FIG_MX1}). It is a consequence of differences in 
angular velocities of null-surfaces with equivalent radiative boundary conditions, ingoing on the 
event horizon and outgoing at infinity.

It is instructive to derive same energetic interaction from frame dragging by inspection 
of Faraday's equation in the Boyer-Lindquist coordinates $(t,r,\theta,\phi)$ of the Kerr 
metric with Killing vectors $k^b=(\partial_t)^b$ and $m^b=(\partial_\phi)^b$. Let \cite{lic67} 
\begin{eqnarray}
{\bf F} = {\bf u}\wedge {\bf e} + * {\bf u} \wedge {\bf h}
\end{eqnarray}
denote the four-vector representation $(u^b,e^b,h^b)$ of the electromagnetic field 2-form
$F_{ab}$, associated with a timelike unit tangent $u^b$, $u^cu_c=-1$, of zero-angular momentum 
observers. Following \cite{bar72,tho82,tho86}, we note the one-form ${\bf u} = -\alpha {\bf d}t$ 
with redshift $\alpha$, whose vector field ${\bf u}=\alpha^{-1}({\bf k}+\omega {\bf m})$
is linear combination of the Killing vectors. Consequently, $\nabla_cu^c=0$. The zero-angular
momentum observers measure an electric field $e^b$ and a magnetic field $h^b$, 
$e^b=u_cF^{ac}$ and $h^b=u_c*F^{cb},$ each of which has three degrees of freedom in view 
of the algebraic constraints $u^ce_c=u^ch_c=0$. In the frame of the observers, we have 
${\bf e}=(0,E^i)$ and ${\bf h}=(0,B^i)$, where $i=1,2,3$ refers to the coordinates of the 
surfaces of constant $t$. The star $*$ denotes the Hodge dual, satisfying $*^2=-1$ in four dimensions. 

We evaluate Faraday's equation 
\begin{eqnarray}
\nabla_a*F^{ab}=0,
\end{eqnarray}
by first considering the expression
$\nabla_a (u^ah^b-u^bh^a) = {\cal L}_u h^b + (\nabla_cu^c)h^b-(\nabla_ch^c)u^b,$
where ${\cal L}_u h^b = (u^c\nabla_c)h^b - (h^c\nabla_c)u^b$ denotes the Lie-derivative 
of $h^b$ with respect to the vector field $u^b$. Projected onto surfaces of constant $t$
(orthogonal to $u^b$), we have
\begin{eqnarray}
\left({\cal L}_{\bf u} {\bf h}\right)_\perp = \alpha^{-1}\left(\partial_t {\bf B} + {\cal L}_\omega {\bf B}\right)
\end{eqnarray}
when evaluated in the frame of the zero-angular momentum observers, where $L_\omega$ 
refers to the Lie-derivative with respect to $\omega^i\equiv\omega m^i$ 
(where $m^i$ is not a unit three-vector). To proceed, we write 
$\nabla_a = D_a- u_a(u^c\nabla_c),$
and note $(*{\bf u}\wedge {\bf h})_{abcd} = \epsilon_{abcd}u^ce^d$, the acceleration 
$(u^c\nabla_c)u_b=\alpha^{-1}\nabla_b\alpha$, and consider
$\nabla^b(\epsilon_{abcd}u^ce^d)=\epsilon_{abcd}(D^bu^c)e^d-\epsilon_{abcd}u^ba^ce^d
+\epsilon_{abcd}u^c\nabla^be^d.$ The projection of the right hand side onto the space coordinates
$i=(r,\theta,\phi)$ normal to $u^b$ satisfies
\begin{eqnarray}
\epsilon_{ibcd}(D^bu^c)e^d+\tilde{\epsilon}_{ijk}a^je^k+\tilde{\epsilon}_{ijk}\nabla^je^k=
\epsilon_{ibcd}(D^bu^c)e^d+\alpha^{-1}\tilde{\epsilon}_{ijk}\nabla^j(\alpha e^k),
\end{eqnarray}
where $\epsilon_{aijk}u^a=\tilde{\epsilon}_{ijk}=\sqrt{h}\Delta_{ijk}$ with
$\sqrt{-g}=\alpha\sqrt{h}$ over the three-volume $\sqrt{h}$ of the spacelike coordinates,
using the fully anti-symmetric symbol $\Delta_{ijk}$, $\Delta_{123}=1$. Here,
the first term on the right hand side vanishes, since $D_bu_c$ is spacelike:
$u^b(D_bu_c)=0$ by construction and $u^cD_bu_c=0$ in view of $u^2=-1$. We conclude that
Faraday's law includes an additional term (derived alternatively in \cite{tho86} and 
references therein)
\begin{eqnarray}
\tilde{\nabla}\times \alpha {\bf E} = -\partial_t {\bf B} + 4\pi {\cal J}_m,
\label{EQN_FAR}
\end{eqnarray}
where $\tilde{\nabla}_i=D_i$ and
\begin{eqnarray}
{\cal J}_m = -\frac{1}{4\pi} {\cal L}_\omega {\bf B}
\label{EQN_JM}
\end{eqnarray}
expressed as if a current of virtual magnetic monopoles. 
Applied to an axisymmetric inner torus magnetosphere, (\ref{EQN_JM}) satisfies
\begin{eqnarray}
\omega_i {\cal J}_m^i \simeq 
 \frac{1}{8\pi} {\bf B}\cdot \tilde{\nabla}(\omega_i\omega^i) > 0,~~
\omega_i\omega^i = 4\frac{z^2\sin^2\lambda}{(z^2+\sin^2\lambda)^3} ~(\theta=\frac{\pi}{2}),
\label{EQN_LOM}
\end{eqnarray}
where the inequality refers to a poloidally ingoing magnetic field 
as sketched in Fig. \ref{FIG_MX1} with the orientation and sign of ${\cal J}_m$ 
as indicated, and $z=r/M$.

By (\ref{EQN_LOM}), frame dragging induced poloidal current loops in the inner torus magnetosphere 
produce Maxwell stresses which tend to spin down a rapidly spinning black hole against 
surrounding matter. In this process, the horizon surface serves as a passive load while 
the surrounding matter is subject to competing torques acting on the inner and the outer 
face of the torus \cite{van99}. The latter should give rise to forced turbulence with 
some similarity to Taylor-Couette flows \cite{ste09}.

\section{Non-thermal high-energy emissions from Kerr black holes}

Here, we focus on frame dragging powering {\em non-thermal} radiation 
processes, even when not necessarily dominant in the total output 
generated by the black hole.

\subsection{High-energy hadronic emissions upstream an Alfv\'en front}

\begin{figure}
\centerline{\includegraphics[scale=.28]{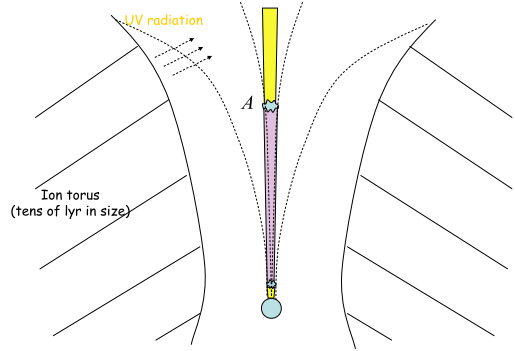}}
\caption{The formation of a linear accelerator for cosmic rays in the funnel of
an extended ion torus powered by an outgoing Alfv\'en front $A$, enabling the 
production of cosmic rays by acceleration of UV-ionized contaminants to GZK energies. 
({Reprinted from} \cite{van09}. (c)2009 The Royal Astronomical Society.)}
\label{FIG_CAP}
\end{figure}

The frame dragging induced potential energy $E=\omega J_p$ extracts leptons of charge $e$ 
with angular momentum \cite{van00}
\begin{eqnarray}
J_p=eA_\phi,~~\Phi=2\pi A_\phi,
\end{eqnarray}
along open magnetic flux-tubes with magnetic flux $\Phi$ out to infinity powered by
a black hole in equilibrium (\ref{EQN_MUH}). 
This {\em capillary effect} \cite{van09} produces an {\em ab initio} leptonic outflow 
carrying a Poynting flux in an approximately force-free state as envisioned in \cite{bla77} 
up to a terminal Alfv\'en front when emanating from time-dependent source, as 
sketched in Fig. \ref{FIG_CAP}. The Alfv\'en front mediates the raw Faraday induced 
horizon potential out to large distances, powering a linear accelerator
upstream \cite{van08a}.

Away from the immediate radiation field of an accretion disk, the linear accelerator
creates UHECRs from ionic contaminants in a UV-irradiated funnel of 
the extended ion torus in an AGN \cite{van09}, where the opacity for UHECRs against 
photon-pion absorption of UHECRs is minimal and curvature radiation drag is negligible.
It meets the requirements for producing UHECRs at GZK energies \cite{wax04} by
intermittent nuclear activity with otherwise limited activity in the mean 
\cite{far08} whenever the duty cycle is small \cite{van09}. Fig.(\ref{FIG_CAP}) illustrates pick-up of ionic 
contaminants by UV irradiation coming off an ion torus, whose linear size is 
light years (e.g. \cite{for95}).

In dimensionful form, the frame dragging induced potential energy (\ref{EQN_E1}) gives
\cite{van09}
\begin{eqnarray}
{\cal E} = 5.6\times 10^{19} \sqrt{\frac{M_9}{T_7}} \left(\frac{\theta_H}{0.5}\right)^2 \mbox{~eV}
\label{EQN_CUH1}
\end{eqnarray}
where a canonical value of 60$^o$ is used for the jet opening angle (as in M87, \cite{jun99}). 
In (\ref{EQN_CUH1}), we have expressed the underlying poloidal magnetic field-energy -- 
difficult to observe directly -- with the finite lifetime of rapid spin of the black 
hole using the correlation (\ref{EQN_B5M9}). Here, $T_7$ is associated with the 
lifetime of the AGN, which is typically on the order of ten million years
(e.g., \cite{ode09}).

The correlation (\ref{EQN_CUH1}) applies to intermittent activity with low duty-cycle of the 
central black hole. Intermittencies in an inner disk, as in MCG 60-30-15 \cite{tan95,iwa96}, 
inevitably modulate the half-opening angle $\theta_H(t)$ in time and, thereby, the 
instantaneous luminosity in the jet \cite{van09}
\begin{eqnarray}
L_j(t)\simeq 1.3\times 10^{46}\left(\frac{M_9}{T_7}\right) \left(\frac{\theta_H(t)}{0.5}\right)^4~\mbox{erg~s}^{-1}
\label{EQN_CUH2}
\end{eqnarray}
up to several orders of magnitude. Thus, energies (\ref{EQN_CUH1}) can be produced
by AGN with low luminosities on average, i.e.,
\begin{eqnarray}
<\theta_H^4(t)><<1
\label{EQN_TH4}
\end{eqnarray}
as in the Seyfert galaxies and Low Ionization Nuclear Emission-line Regions (LINERs)
associated with the PAO results \cite{zaw08}.

\subsection{UHECRs from rotating black holes in low luminosity AGN}

\begin{figure}
\centerline{\includegraphics[scale=0.40]{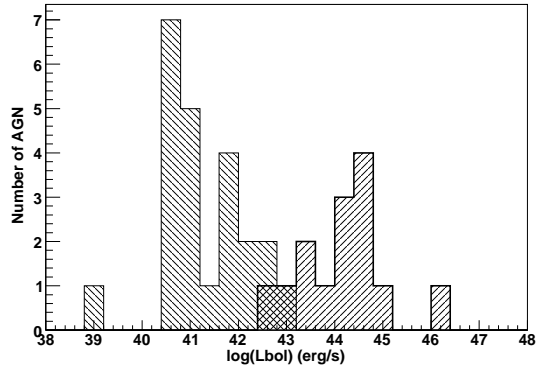}}
\caption{The distributions of galaxy luminosities in the Virgo cluster (left)
and the luminosities of Seyferts and LINERs within the 3.2 degree 
circle of UHECRs detected by the PAO with energies above 5.7 EeV (right) are
distinct and nearly disjoint, where the latter represent low luminosity AGN between 
a few times $10^{42}$ erg s$^{-1}$ and $10^{46}$ erg $s^{-1}$.
({Reprinted with permission from} \cite{zaw08}. (c)2009 The Americal Astronomical Society.)}
 \label{FIG_ZAW}
 \end{figure}
 
The Pierre Auger Observatory (PAO) identified an angular correlation of UHECR 
with the matter distribution in the Local Universe traced by nearby AGN in a sample 
of 27 UHECRs with energies above the GZK threshold. This result was recently extended 
to a sample of 58 events \cite{pao09}, although the larger sample size shows no 
improvement of the correlation to nearby AGN.

The initial PAO data suggest a tentative UHECR association with spiral galaxies based on 
the all Sky Survey (HIPASS) \cite{ghi08}. Extended with the NASA/IPAC Extragalactic 
Database (NED), the results point to low-luminosity Seyfert galaxies and Low-Ionization 
Nuclear Emission Line Regions (LINERs) with very few radio galaxies \cite{mos08}, except 
for apparent multiple events coming from Cen A \cite{pao09}. There is no indication of 
BL Lac objects \cite{har07} and there appears to exist a minimum bolometric luminosity for 
UHECRs activity, consistent with the paucity of UHECRs from the Virgo cluster \cite{zaw08,pao09}. 
An association with radio-galaxies \cite{nag08} might reflect diffusive acceleration 
of charged particles in radio lobes \cite{fra08b}, but this would require the extended 
radio-jet to be baryon-rich and is at odds with the above-mentioned 
Seyfert galaxies and LINERs \cite{mos08}.

A low average jet luminosity $<L_j>$ in (\ref{EQN_CUH2}) by a small time average 
of the horizon half-opening angle (\ref{EQN_TH4}) is consistent with the observed 
statistics on the association of 27 UHECR-AGN, compiled in Fig. \ref{FIG_ZAW}
by \cite{zaw08}
and intermittent behavior in the Seyfert nucleus shown in Fig. \ref{FIG_MCG}.
The proposed UHECR-AGN association predicts that the PAO should detect repeat events 
from some of its source directions. 

While a rigorous identification of nearby sources of UHECR remains elusive, AGN 
remain important candidates while GRBs are essentially ruled out on the basis of 
their low event rates \cite{van09}.

\subsection{High-energy photon emissions downstream an Alfv\'en front}

Downstream of the Alfv\'en front terminating the capillary outflow, 
intermittencies in the nucleus inevitably produce shocks by steepening, giving
rise to non-thermal emissions by dissipation of kinetic energy.
Capillary jets are non-uniform in luminosity and variability within their observed
half-opening angle $\theta_j$ (generally distinct from $\theta_H$ by collimation). 
Flux surfaces carry maximal magnetic flux at the interface with the environment or 
a surrounding baryon-rich disk or torus wind. As such, surfaces of maximal flux at 
$\theta\simeq \theta_j$ should have
\begin{itemize}
\item Maximum luminosity, by the frame dragging induced Faraday potential energies; 
\item Maximum variability, in a turbulent boundary layer between the jet and a
      collimating wind.
\end{itemize}
It accounts for a positive correlation between luminosity and variability \cite{rei01,loy02}, 
since the viewing angle selects a small range of magnetic flux surfaces by relativistic 
beaming: close on-axis, we see a smooth light curve of moderate luminosity, while off 
axis (within $\theta_j$), we see a luminous and variable light curve produced by outer 
flux surfaces.

\subsection{GRBs from rotating black holes}

\begin{figure}
\centerline{\includegraphics[width=60mm,height=40mm]{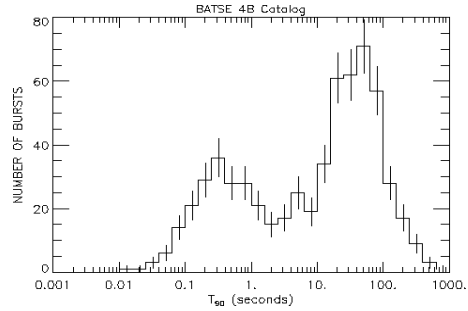}}
\caption{The bi-modal distribution of durations $T_{90}$ of GRBs around
0.3 s and 30 s reveals the existence of at least two classes of events. 
For GRBs from rotating black holes, the long and short durations agree 
quantitatively with hyperaccretion onto slowly rotating black holes and 
suspended accretion around rapidly rotating black holes, respectively. GRB 
emissions from black hole spin hereby predicts (weak) X-ray afterglows also 
to short bursts, confirmed by HETE II and {\em Swift} in 2005. (Courtesy of NASA 
MSFC, Space Sciences Laboratory.)}
\label{FIG_BIM}
\end{figure}

The BATSE catalogue reveals a bi-modal distribution of durations, centered around
0.3 s and 30 s \cite{kou93}, possibly with a third distribution in between (Fig. \ref{FIG_BIM}).
HETE II and {\em Swift} observations reveal long GRBs with and without supernovae, and a
diversity in their X-ray afterglow emissions. 
This diversity is consistent with a GRBs as a common endpoint of various 
astronomical scenarios, such as those in Fig. \ref{FIG_MX0}. Black holes 
formed in core-collapse supernovae in binaries should be rapidly rotating 
\cite{bar70,pac98,van04}.

The durations of tens of seconds of long GRBs should be viewed as intrinsic 
to a long-lived inner engine, as inferred from short timescale variabilities
\cite{pir98}. The inferred ultra-relativistic baryon-poor jets powering the GRBs 
should be viewed in direct connection to the event horizon of the black hole
\cite{lev93}, here by an open flux tubes supported by (\ref{EQN_MUH}) whereby
gamma-ray emissions are geometrically beamed. It leads to true energies in 
gamma-rays with a relatively narrow distribution around $10^{51}$ erg \cite{fra01}
-- a notably {\em small} fraction of the energy reservoir of a rapidly rotating 
black hole \cite{van03}.

An angular momentum powered inner engine is consistent with highly aspherical 
supernovae \cite{hoe99}. It is here powered by aspherical irradiation 
{\em from within} by magnetic disk winds \cite{van03}, and forms a specific realization
of \cite{bis70}. The small fraction of collapsars producing GRB-supernovae may 
result from the low probability of leaving the newly formed black hole centered, 
as in the presence of the Bekenstein gravitational-radiation recoil mechanism, 
consistent with the observed branching ratio of about 0.2-0.4\% of the SN Type Ib/c 
producing GRBs \cite{van04}.

The bi-modal distribution in durations is quantitatively consistent with GRBs from 
{\em slowly} and {\em rapidly} spinning black holes in a state of hyper- and, respectively, 
suspended accretion \cite{van01}, where the latter is illustrated in Fig. \ref{FIG_MX0}.

\subsubsection{A spectral-energy correlation in capillary jets including $T_{90}$}

As discussed above, the GRB-afterglow emissions represent the dissipation of kinetic energy
in the capillary outflow, here in internal and external shocks \cite{she90}. Let $c_1$ denote 
the ratio of observed peak energy $E_p$ to the {\em ab initio} potential energy ${\cal E}$ in 
(\ref{EQN_E1}). It represents converting ${\cal E}$ into kinetic energy in a neutron-enriched 
leptonic jet with subsequent conversion into high-energy gamma-rays in shocks. The luminosity 
in Poynting flux along the open magnetic flux-tube in the force-free limit \cite{gol69,bla77,tho86} 
is $L=\Omega^2_AA_\phi^2,$ where $\Omega_A$ denotes the angular velocity of the Alfv\'en front.
Let $c_2$ denote the efficiency of converting $L$ into the true energy in $\gamma$-rays: 
$E_{\gamma}=c_2LT_{90}$. Then (\ref{EQN_E1}) predicts $E_pT_{90}^{1/2} = ek E_{\gamma}^{1/2}$, 
where $k=2c_1/\sqrt{c_2}$. Here, we identify $T_{90}$ with the lifetime of spin of an initially
rapidly rotating black hole, as discussed in detail in \S7 below. 

In the approximation of commensurate angular velocities, i.e., 
$\Omega_A$ to be of order $\Omega_H$, the Ghirlanda relation \cite{ghi04} 
$E_p\propto E_\gamma^{0.7}$, the Eichler \& Jontof-Hutter correlation between $E_p$ and kinetic 
energy of the outflow \cite{eic05}, implies $c_2\propto E_p^{3/2}$. Following a heuristic 
{\em Ansatz} $c_2\propto c_1$, $k\propto E_{\gamma}^{21/40}\simeq E_{\gamma}^{1/2}$ and hence 
\begin{eqnarray}
E_pT_{90}^{1/2} \propto E_{\gamma},
\label{EQN_C1}
\end{eqnarray}
where $E_\gamma$ shows aforementioned narrow distribution around $10^{51}$ erg \cite{fra01}). 
\begin{center}
\begin{figure}
\center{\includegraphics[angle=00,scale=.33]{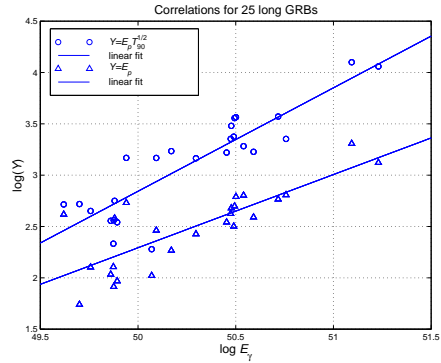}}
\caption{Comparison of a spectral-energy correlation (\ref{EQN_C1}) for GRBs from
spin-down of rotating black holes against HETE II and {\em Swift} data.  A linear fit
gives a slope and Pearson coefficient $(s, c) = (1.01, 0.85)$ to 
$Y = \log \left(E_pT^{1/2}\right)$ ($circles$) and $(s, c) = (0.71, 0.76)$ in 
the Ghirlanda correlation to $Y = \log (E_p)$ ($triangles$).
({Reprinted from} \cite{van08a}. (c)2008 The American Astronomical Society.)}
\label{FIG_C1}
\end{figure}
\end{center}
Fig.\ref{FIG_C1} compares (\ref{EQN_C1}) to a compilation of HETE II and {\em Swift} data 
\cite{ama02,ghi04,ghi07}. See also \cite{eic04,ghi04} for an alternative perspective on the 
basis of viewing angles.

\section{Evolution of Kerr black holes surrounded by a torus magnetosphere}

With no small geometrical parameter present in the inner torus magnetosphere,
we can neglecting the sub-dominant energetic output along the spin-axis. The
evolution of the black hole is then dominated by (\ref{EQN_LOM}), giving rise
to the {\em closed model approximation} given by a torque $T$ and black hole 
luminosity $L_H$ in the inner torus magnetosphere satisfying \cite{van99}
\begin{eqnarray}
T=-\dot{J},~~L_H=-\dot{M},
\label{EQN_EV1}
\end{eqnarray}
whereby the torus assumes a state of suspended accretion \cite{van01}. For
a black hole in its lowest energy state (\ref{EQN_MUH}), (\ref{EQN_EV1})
determines its lifetime $T_{s}$ of rapid spin. To calculate this, we next derive 
some relations between the relevant parameters.

The closed model approximation follows in the absence of a small parameter in the
poloidal cross section of the torus magnetosphere, connecting the event horizon
of the black hole and the surrounding matter. As inner engines to GRBs, neglecting 
the energy output in high energy emissions about the spin axis of the black hole is 
consistent with the observed small true energy in gamma-rays of about $10^{51}$ erg 
\cite{fra01}, relative to the spin energy of about $10^{54}$ erg of a stellar mass 
black hole.

\subsection{Relations between $B$, $M$ and $T_{s}$}
 
The frame dragging induced spin down of a rapidly rotating black hole against surrounding matter
give rise to a viscous timescale for the lifetime of black hole spin,
$T_{s} =T_{7}\times 10^7\mbox{~s}\simeq 29\% \times 2\times 10^{63} M_9 \mbox{~erg~} D_H^{-1}$,
where $D_H=\frac{c}{12}B^2M^2=5.6\times 10^{47}\left(B_5M_9\right)^2\mbox{erg s}^{-1}$ denotes
the horizon dissipation rate in terms of $B=B_5\times 10^5$ G and $M=M_9 \times 10^9$ $M_\odot$
\cite{van09}. It defines a correlation $B_5M_9^{1/2}T_{7}^{1/2}\simeq 1.04$, or \cite{van09}
\begin{eqnarray}
B_5M_9 = 1.04 \sqrt{\frac{M_9}{T_{7}}}.
\label{EQN_B5M9}
\end{eqnarray}
Scaled to stellar mass black holes surrounded by superstrong magnetic fields, 
$D_H=6.9\times 10^{52} B_{15}^2\left(\frac{M}{7M_\odot}\right)^2 \mbox{~erg~s}^{-1}$, 
we find
\begin{eqnarray}
\left(\frac{B}{5\times 10^{15}\mbox{G}}\right)\left(\frac{M_\odot}{7M_\odot}\right)
=1.05 \sqrt{\frac{M~20\mbox{~s}}{7M_\odot T_{s}}},
\label{EQN_B5M7}
\end{eqnarray}
showing a characteristic lifetime $T_{s}$ of spin of tens of seconds.

\subsection{Relations between $R_T$, $M_T$ and $T_{s}$}

The energy ${\cal E}_B$ in net poloidal magnetic flux that a torus with kinetic energy
${\cal E}_k$ can support is 
subject to a stability bound \cite{van03} 
\begin{eqnarray}
\frac{{\cal E}_B}{{\cal E}_k}\le \frac{1}{15}.
\label{EQN_BSTAB}
\end{eqnarray}
At the stability bound (\ref{EQN_BSTAB}), the lifetime of spin and the mass $M_T$ of the 
torus satisfy \cite{van09}
\begin{eqnarray}
M_T\simeq 120 M_\odot \left(\frac{{\cal E}_k}{15{\cal E}_B}\right)
\left(\frac{R_T}{6R_g}\right)^4 \left(\frac{M_9^2}{T_7}\right),
\label{EQN_M1}
\end{eqnarray}
\begin{eqnarray}
M_T\simeq 0.1M_\odot \left(\frac{{\cal E}_k}{15{\cal E}_B}\right) 
\left(\frac{R_T}{6R_g}\right)^4 \left(\frac{M}{7M_\odot}\right)^2
\left(\frac{20\mbox{s}}{T_s}\right)
\label{EQN_M2}
\end{eqnarray}
with characteristic densities of $7.9\times 10^{-11}$ g cm$^{-3}$ and, respectively, 
$1.9\times 10^{11}$ g cm$^{-3}$ (close to the neutron drip line). The associated
sound velocities at canonical temperatures of 10 keV and 2 MeV, respectively, are
\begin{eqnarray}
c_s=0.0041c,~~c_s=0.0516c,
\label{EQN_CS1}
\end{eqnarray}
where $c$ denotes the velocity of light. The strength of the (poloidal) magnetic field in 
(\ref{EQN_M1}-\ref{EQN_M2}) can be conveniently expressed in terms of the associated 
Alfv\'en velocities (\ref{EQN_VA}) according to the respective estimates
\begin{eqnarray}
v_A=0.1052c,~~v_A=0.1072c.
\label{EQN_VA1}
\end{eqnarray}

\subsection{A system of evolution equations}

The closed model approximation (\ref{EQN_EV1}) gives evolution equations
\begin{eqnarray}
\left\{\begin{array}{cl}
\dot{M}&=-e\Omega_H\left(1-\eta\right)\Omega_T\\
\dot{J}&=-e\Omega_H\left(1-\eta\right)
\end{array}\right.,
\label{EQN_EVO}
\end{eqnarray}
where 
\begin{eqnarray}
\eta = \frac{\Omega_T}{\Omega_H}
\label{EQN_ETA}
\end{eqnarray}
denotes the angular velocity of the torus relative to $\Omega_H$ and $e$ 
represents the strength of inner torus magnetosphere, that can be solved by 
numerical integration with {\em closure relations} for $\Omega_T$. We shall 
consider the two alternatives
\begin{eqnarray}
A:~~\Omega_T=\Omega_{ISCO}
\label{EQN_MA}
\end{eqnarray}
and
\begin{eqnarray}
B:~~\Omega_T=\frac{1}{2}\Omega_H.
\label{EQN_MB}
\end{eqnarray}
The first is motivated by observational results shown in Fig. \ref{FIG_MCG}.
The second represents balance between angular momentum flux from the black 
hole and to infinity in magnetic torus winds for a symmetric distribution 
of poloidal magnetic flux between the inner and outer torus magnetospheres.

For an initially maximally spinning black hole, numerical integration of (\ref{EQN_EVO}) 
shows an overall efficiency of close to 60\% (equal to when $a/M=0.8$ initially) for 
(\ref{EQN_MA}), and an efficiency of 35\% for (\ref{EQN_MB}). In light of these
efficiences, spin down of a black hole against surrounding matter is largely
a viscous process, representing the creation of astronomical amounts of 
Bekenstein-Hawking entropy. 

We note that integrations of modified equations of suspended accretion 
that include precession (with no gravitational radiation) have also been 
performed \cite{lei07}.

\subsection{Observing spin down in a normalized light curve of 600 long GRBs}

We apply matched filtering to two sets of 300 long GRBs from the BATSE catalogue,
each normalized in durations and count rates by application of matched filtering
\cite{van09}. 
The resulting normalized light curves (nLC) can be directly summed for averaging.
In this process, short and intermediate timescale fluctuations or modulations are 
filtered out. We also ignore baseline levels in background gamma-ray count rates, 
by matching templates to data up to an arbitrary offset in count rate.

Our templates are calculated by numerical integration of (\ref{EQN_EVO}) with 
closure (\ref{EQN_MA}) or (\ref{EQN_MB}), governing the leading-order evolution 
of the black hole, here considered with initially maximal spin $(\lambda=\pi/2)$.
\begin{center}
\begin{figure}
\center{\includegraphics[angle=00,scale=.33]{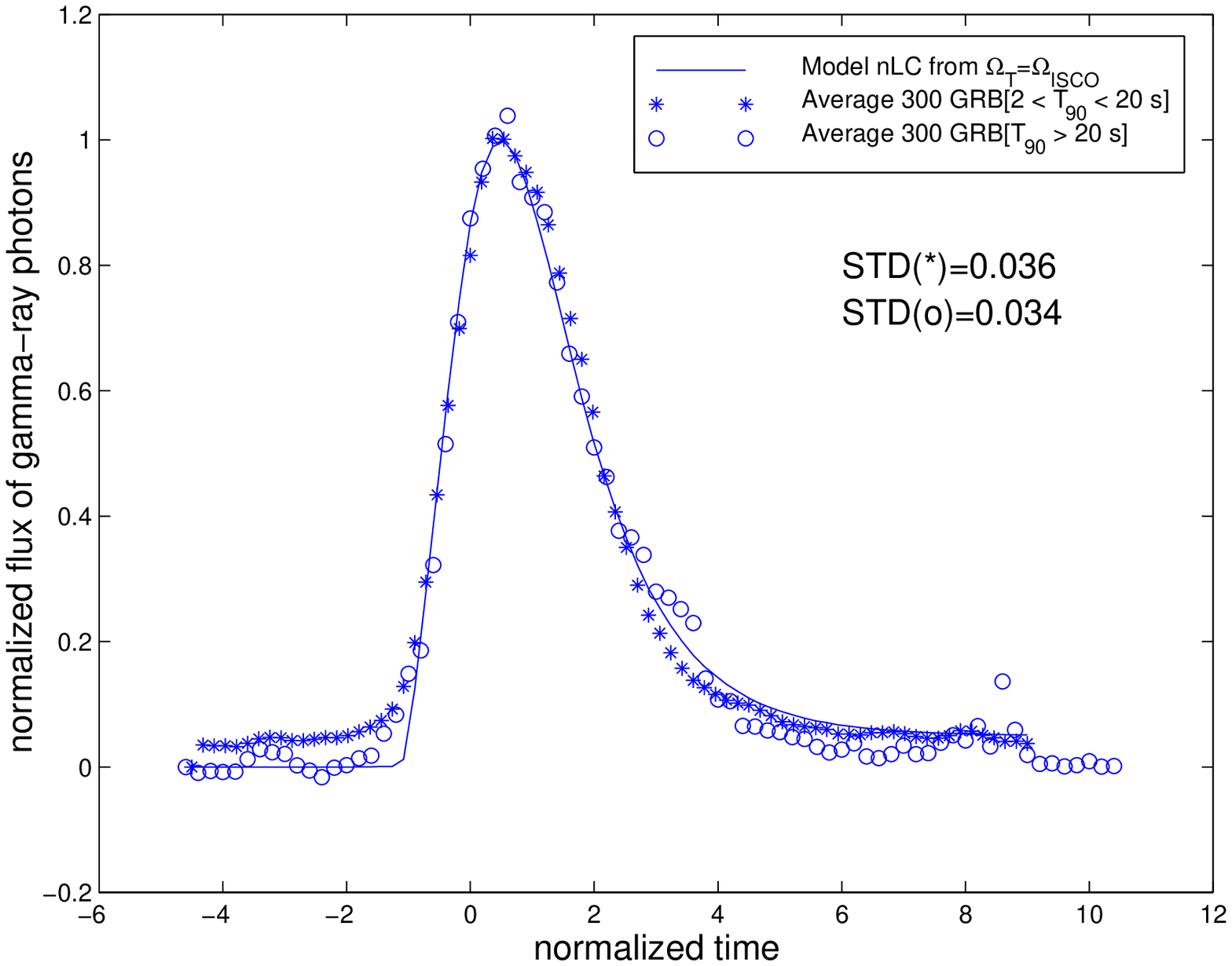}
        \includegraphics[angle=00,scale=.33]{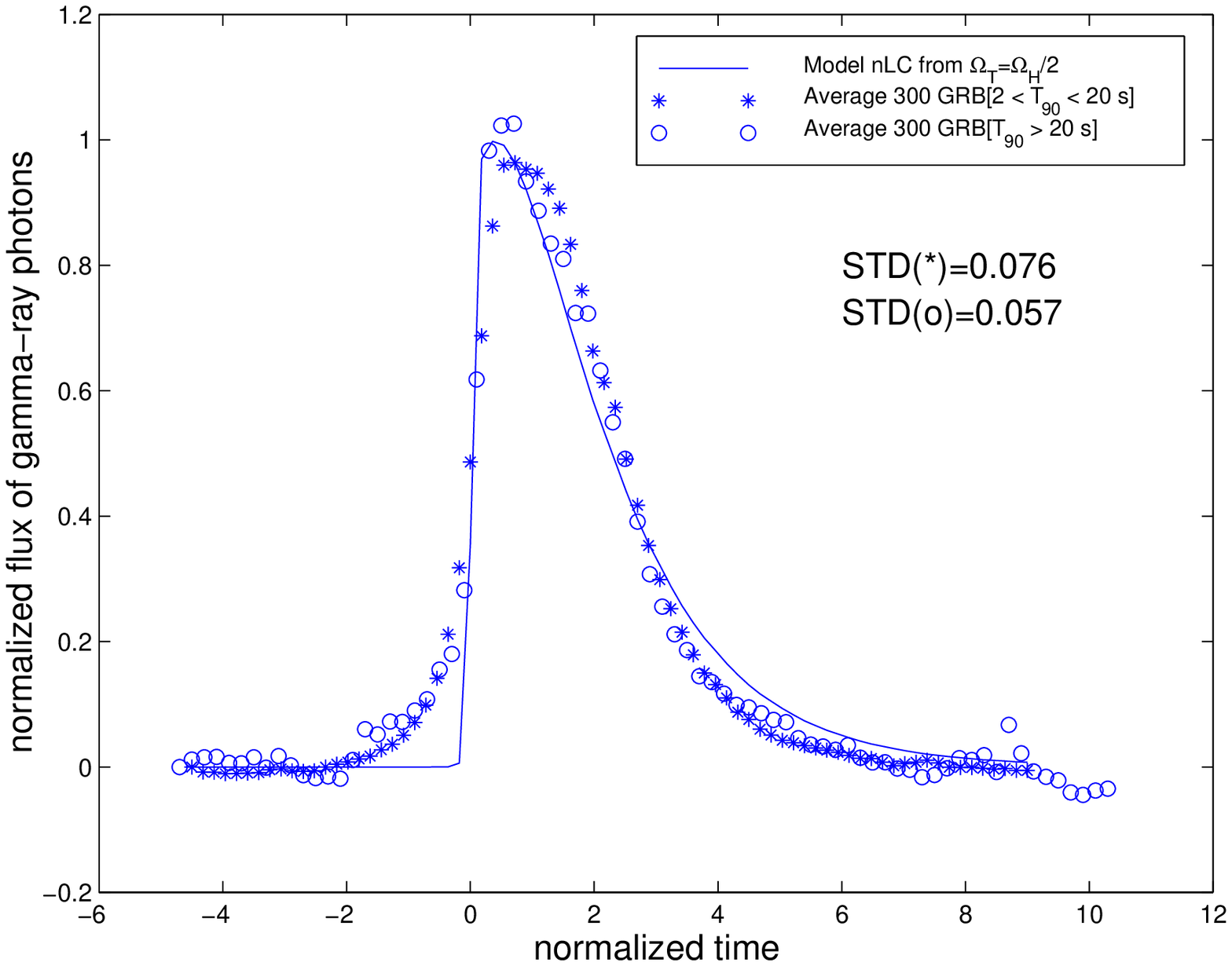}}
\caption{Shown are normalized light curves (nLC) of long GRBs. The nLCs are produced by 
averaging sets of 300 individually normalized light curves by application of matched 
filtering against two model templates for viscous spin-down of rapidly rotating black holes 
by a torus. The sets are 300 bursts of 2 s$<T_{90}<$20 s and 300 bursts of $T_{90}>20$ s in
the BATSE Catalogue. Shown are results for tori with angular velocities $\Omega_T=\Omega_{ISCO}$ 
($left$, {reprinted from} \cite{van09}. (c)2009 The Royal Astronomical Society.) 
and $\Omega_T=\frac{1}{2}\Omega_H$ ($right$). The residual standard deviations between 
the generated nLC and the model template indicate a preference for the first alternative.}
\label{FIG_nLC}
\end{figure}
\end{center}
The minor energy output in gamma-rays are modeled subsequently for the calculated
evolution of the black hole, by further positing a positive correlation between 
the radius $R_T$ of the torus and $\theta_H$ \cite{van09}, i.e.: 
\begin{eqnarray}
L_j\propto \Omega_H^2 \theta_H^4 {\cal E}_B,~~
{\cal E}_T \simeq \frac{1}{2}e(z)\Omega_T^2R_T^2,
\label{EQN_LJ}
\end{eqnarray}
where $e(z)$ is given in (\ref{EQN_EL}), $z=R_T/M$, $\Omega_T$ in (\ref{EQN_OMT}), 
${\cal E}_B\propto {\cal E}_T$ by (\ref{EQN_BSTAB}), using 
\begin{eqnarray}
\theta_H\propto z^{1/2}
\end{eqnarray} 

With closure (\ref{EQN_MA}) $L_j(t)$ starts at a finite value $L_j(t_0)>0$ at the
time of onset $t_0$ for an initially maximally spinning black hole, and gradually 
increases to a maximum before decaying to a finite value as $\Omega_H$ approaches 
$\Omega_{ISCO}$. The maximum is attained with a delay
\begin{eqnarray}
\frac{\tau}{T_{90}}\simeq 16\% ~(25\%) 
\label{EQN_tau}
\end{eqnarray}
relative to $T_{90}$ of the model burst $L_j(t)-L_j(t_0)\ge0$ ($L_j(t)$), where 
$t_0$ denotes the time of onset of the burst in template (\ref{EQN_MA}) and 
max $L_j(t)/L_j(t_0)=3.27$. In the second case, (\ref{EQN_LJ}) with (\ref{EQN_MB}), 
$L_j(t)$ starts promptly at near-maximum, and rapidly decays with black hole spin. 

Fig. \ref{FIG_nLC} shows the result for models (\ref{EQN_MA}-\ref{EQN_MB}). 
The nLC generated by the template of A (\ref{EQN_MA}) is remarkably consistent with 
the template itself, and especially so for the average of the 300 GRBs with 
$T_{90}>$20 s \cite{van09}. Very similar results are obtained for long bursts in the 
current {\em Swift} catalogue. The result is numerically superior to the nLC generated 
by the template of B (\ref{EQN_MB}), as seen by comparing the standard deviations 
of residuals relative to the model light curves,
\begin{eqnarray}
\sigma_A = 0.036~~(300\mbox{~GRBs}[2<T_{90}<20]),~0.034~~(300\mbox{~GRBs}[20<T_{90}])
\label{EQN_SA}
\end{eqnarray}
and
\begin{eqnarray}
\sigma_B = 0.076~~(300\mbox{~GRBs}[2<T_{90}<20]),~0.057~~(300\mbox{~GRBs}[20<T_{90}]),
\label{EQN_SB}
\end{eqnarray}
as shown in Fig. \ref{FIG_nLC}. 

We conclude that long GRBs are spin powered, not accretion powered \cite{van09} 
based on GRB060614 and our matched filtering results:
\begin{enumerate}
\item
GRB060614 without a supernova was powered by a naked inner engine produced in a 
merger event with a duration $T_{90}=102$ s. No such long time scale exists in 
hyper-accretion in a merger event and, instead, it calls for a secular 
time scale, here identified with the viscous time scale of tens of 
seconds (100 s for superstrong magnetic fields of $B=10^{16}$ G in \cite{van99}) 
for the lifetime of black hole spin;
\item
Black hole spin down against matter at the ISCO induced by (\ref{EQN_LOM}) gives
rise to a model light curve which closely matches the data in Fig. \ref{FIG_nLC},
and forms a specific realization of the direct connection of the observed GRB 
emissions and the event horizon of a black hole \cite{lev93}.
\end{enumerate}

Accretion powered models (e.g \cite{woo93,kum08}) are different. For long bursts, they
apply to core-collapse events, taking into account the free fall timescale of the
stellar envelope, but not to mergers. Disk winds are too contaminated (e.g. \cite{van03}) 
to produce the required baryon-poor outflows for the prompt GRB emissions. If attributed 
to the black hole, spin {\em up} due to accretion \cite{kum08} is at odds with the decay 
in the nLC Fig. \ref{FIG_nLC}. 
 
Without accretion, open outflows producing effective spin down are at odds with the 
true energy in gamma-rays \cite{fra01}, unless efficiencies are absurdly low. 
Furthermore, the angular velocity of the open flux tubes is about $\Omega_H/2$ in view of 
matched impedances of the event horizon of the black hole and infinity \cite{bla77,tho86}. 
But template of B (\ref{EQN_MB}) gives a sub-optimal matching to data (\ref{EQN_SA}-\ref{EQN_SB}), 
showing that open outflows are energetically sub-dominant.

We note that out analysis goes beyond consideration of luminosities alone, which is 
known to be insufficient to identify between these two alternative accretion
modes \cite{liv99}. 

\section{Angular momentum transport in GWs versus MeV-neutrinos}

For stellar mass black holes powering long GRBs, the equations of suspended accretion 
describe catalytic conversion of black hole spin energy and angular momentum in various 
radiation channels, given by \cite{van01,van03}
\begin{eqnarray}
\tau_+ = \tau_- + \tau_{GW}+\tau_\nu,~~\Omega_+\tau_+= \Omega_-\tau_-+\Omega_T\tau_{GW}+P_\nu,
\label{EQN_SA1}
\end{eqnarray}
where $\tau_+\propto(\Omega_H-\Omega_+)$, $\tau_{GW}$ and $\tau_-$ denote the angular momentum
fluxes onto the inner face, in gravitational radiation and in magnetic winds. The latter have
luminosities $L_{GW}=\Omega_T\tau_{GW}$ and $L_w=\Omega_-\tau_-$, respectively. Here, $\Omega_\pm$ denote the
angular velocities of the inner and outer faces of the torus with 
$[\Omega]=\Omega_+-\Omega_-\simeq qb/a=2q\delta$.
The release of angular momentum in neutrinos satisfies $\Omega_T\tau_\nu=P_\nu (v_T/c)^2$, where 
$v_T=\Omega_TR_T$ denotes the angular velocity of the torus and $c$ is the velocity of light.

Solutions to (\ref{EQN_SA1}) exist in which all luminosities scale with the energy in the
magnetic field, for a flat infrared spectrum of the turbulent energy in MHD stresses up to 
the first geometrical break $m^*\simeq a/b$ (\cite{van01,van03}, neglecting $\tau_\nu$). A 
related numerical example on MHD turbulence is given in \cite{wor08}. The associated quadrupole 
mass moment to support the gravitational wave emissions is herein determined self-consistently, 
whose amplitude is consistent with the stability bound (\ref{EQN_BSTAB}).
\begin{figure}
\centerline{\includegraphics[angle=00,scale=.35]{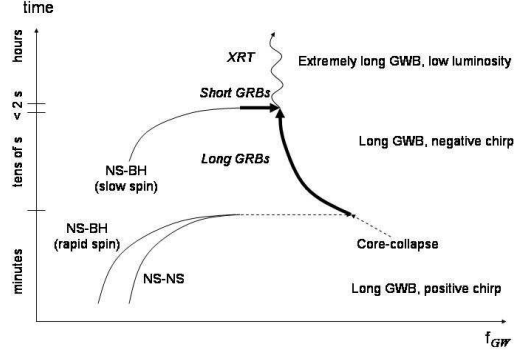}}
\caption{Time-frequecy diagram of gravitational-wave emissions associated with
GRBs from rotating black holes, shown as a function of initial conditions on black 
hole spin emerging out of mergers and core-collapse events. Mergers of neutron stars
with a rapidly (slowly) rotating black hole companion produce long (short) GRBs and 
long (short) bursts in gravitational radiation with a negative chirp, and likewise 
for GRB-supernovae. Mergers produce a preceding positive chirp during binary coalescence. 
The end point of all GRBs is universal: a black hole with slow spin at or about the 
angular velocity of the ISCO. Late time X-ray emissions may be produced for an extended
duration by accretion of remnant matter falling in from an outer envelope \cite{kum08},
or from remnants of a messy break-up of a neutron star \cite{lee98,lee99,ros07,van09}. 
Gravitational-wave emissions accompanying an XRT, if any, are accretion powered,
not spin-powered.}
\label{FIG_GWD}
\end{figure}

For slender tori given by small $\delta$, consider (\ref{EQN_SA1}) in the absence of 
gravitational radiation, i.e., $\tau_{GW}=0$. We then have
\begin{eqnarray}
\frac{\Omega_T}{\Omega_H} = \frac{1}{2}\frac{1}{1+\delta q(\Gamma^2-1)}+O(\delta^2),~~
\frac{P_\nu}{L_w}=2\delta \Gamma^2+O(\delta^2)
\label{EQN_SA2}
\end{eqnarray}
in the limit of a symmetric flux distribution, in which the net poloidal flux in the inner torus 
magnetosphere and in the magnetic winds are equal, where $\Gamma=1/\sqrt{1-v_T^2/c^2}$. 
This result $\Omega_T/\Omega_H\simeq \frac{1}{2}$ in (\ref{EQN_SA2}) is testimony of the 
{\em inefficiency} of angular momentum transport in neutrino flux. 
Fig. \ref{FIG_nLC} shows that it is disfavored by matched filtering in gamma-ray data.

We conclude that $\Omega_T=\Omega_{ISCO}$ is favored by (\ref{EQN_SA}-\ref{EQN_SB}), 
pointing to efficient angular momentum transport in gravitational radiation in (\ref{EQN_SA1}). 
The results for GRBs from rotating black holes are schematically summarized in Fig. \ref{FIG_GWD},
showing histories of GRBs and their emissions in gravitational waves in terms of
distinguishing chirps in gravitational radiation. A {\em negative chirp} in gravitational 
radiation from long bursts is characteristic for being spin powered, as opposed to 
accretion powered in \cite{kob03}. The slowly spinning black hole remnant, common
to all GRBs, may be subject to continuing accretion from remnant matter and debris,
thereby powering a tail in X-ray emissions \cite{lee98,lee99,ros07,van09}, possibly 
accompanied by low-luminosity gravitational radiation.

Similar considerations can be given for (\ref{EQN_SA1}) for tori around supermassive
black holes, upon considering X-ray emissions rather than neutrino emissions, in epochs
when the inner disk reaches down to the ISCO (Fig. \ref{FIG_MCG}). 

\subsection{Pressure driven Papaloizou-Pringle instabilities}

The formation of a non-axisymmetric torus occurs naturally in response to the action
of the black hole onto the surrounding matter. As matter is heated by dissipation of
the energy input from the black hole in turbulent MHD stresses \cite{van99}, it forms 
a {\em hot torus} rather than a disk, and creates a boundary layer between the black hole 
and the extended accretion disk. Its motion is described by a rotation index $q$ in the 
angular velocity distribution $\Omega_T(r)=\Omega_a(a/r)^q$, where $a$ denotes the major 
radius of the torus in a cylindrical coordinate system $(r,\phi)$ in the approximation of 
Newtonian gravity.

Papaloizou-Pringle \cite{pap84} discussed the stability of infinitely slender tori 
in the local linearization $\Omega_T(r)=\Omega_a\left({a}/{r}\right)^q,~~\Omega_a^2={M}/{a^3},$
about a point mass $M$. Here, $q=1.5$ corresponds to Keplerian motion, 
appropriate for cold thin disks, while $q>1.5$ corresponds to pressurized tori up to the Rayleigh 
bound $q=2$, at which the torus becomes unstable against axisymmetric perturbations. Quite generally,
$1.5<q<2$ describes tori with super-Keplerian and sub-Keplerian motion on the inner and,
respectively, outer face. The excess angular momentum, above the Keplerian values, on the
inner face is a driving force towards instability in accord with the Rayleigh criterion.

Tori of finite slenderness $0<\delta=b/(2a)<1/2$, where $b$ denotes the minor radius, 
become instable for $q>\sqrt{3}$ for increasingly many $m$ as $\delta$ decreases \cite{van02}, 
which contains the Papaloizou-Pringle result $q=\sqrt{3}$ in the limit as $\delta=0$. A quadratic 
approximation for the $m=2$ quadrupole waves is 
$q = 1.73 + 0.1 \left({\delta}/{0.1}\right)^2$ \cite{van03}.

The enthalpy $h(r)$ of the torus in the equatorial plane $z=0$ satisfies
$\partial_r h = -M/r^2 + \Omega_T^2 r - \rho^{-1}\partial_r P$, subject to $h=0$ on 
the inner and outer faces of vanishing total pressure, $P=0$, comprising the sum of 
thermal plus magnetic pressures, and $\rho$ denotes the density. For $b<<a$ and in 
the approximation of an incompressible fluid \cite{gol86}, the pressure distribution 
therefore satisfies \cite{van02} ${P}/{\rho} = \left(q-{3}/{2}\right)\Omega_a^2(b^2-x^2),$ 
where $\Omega_a^2=M/a^3$ and $x=r-a$. Thus, $q$ correlates to the central pressure, $P_a$ 
at $r=a$ by $q=1.5+\left({a}/{M}\right)\left({a}/{b}\right)^2{P_a}/{\rho},$ and hence
thermal pressure (neglecting magnetic fields) alone reduces it to \cite{van03}
\begin{eqnarray}
q=1.5+0.2\left(\frac{a}{4M}\right)\left(\frac{\delta}{0.1}\right)^{-2} 
		 L_{\nu,52}^{1/6}\left(\frac{M_T}{0.1M_\odot}\right)^{-1/6}.
\label{EQN_QT}
\end{eqnarray}
associated with a neutrino luminosity $L_{\nu,52}$ for a torus of mass $M_T$, scaled to 
1\% of the mass of a stellar mass black hole. The associated temperatures of
about 2 MeV are noticeably low compared to neutrino energies of up to tens of 
MeV in SN1987A. According to (\ref{EQN_QT}), heating by energetic input from a rotating 
black hole can drive $q$ to a critical point, beyond which the torus becomes unstable 
against any of the non-axisymmetric Papaloizou-Pringle modes. Fig. (\ref{FIG_1201}) shows 
the critical temperature as a function of slenderness ratio.
\begin{center}
\begin{figure}
\center{\includegraphics[angle=00,scale=.25]{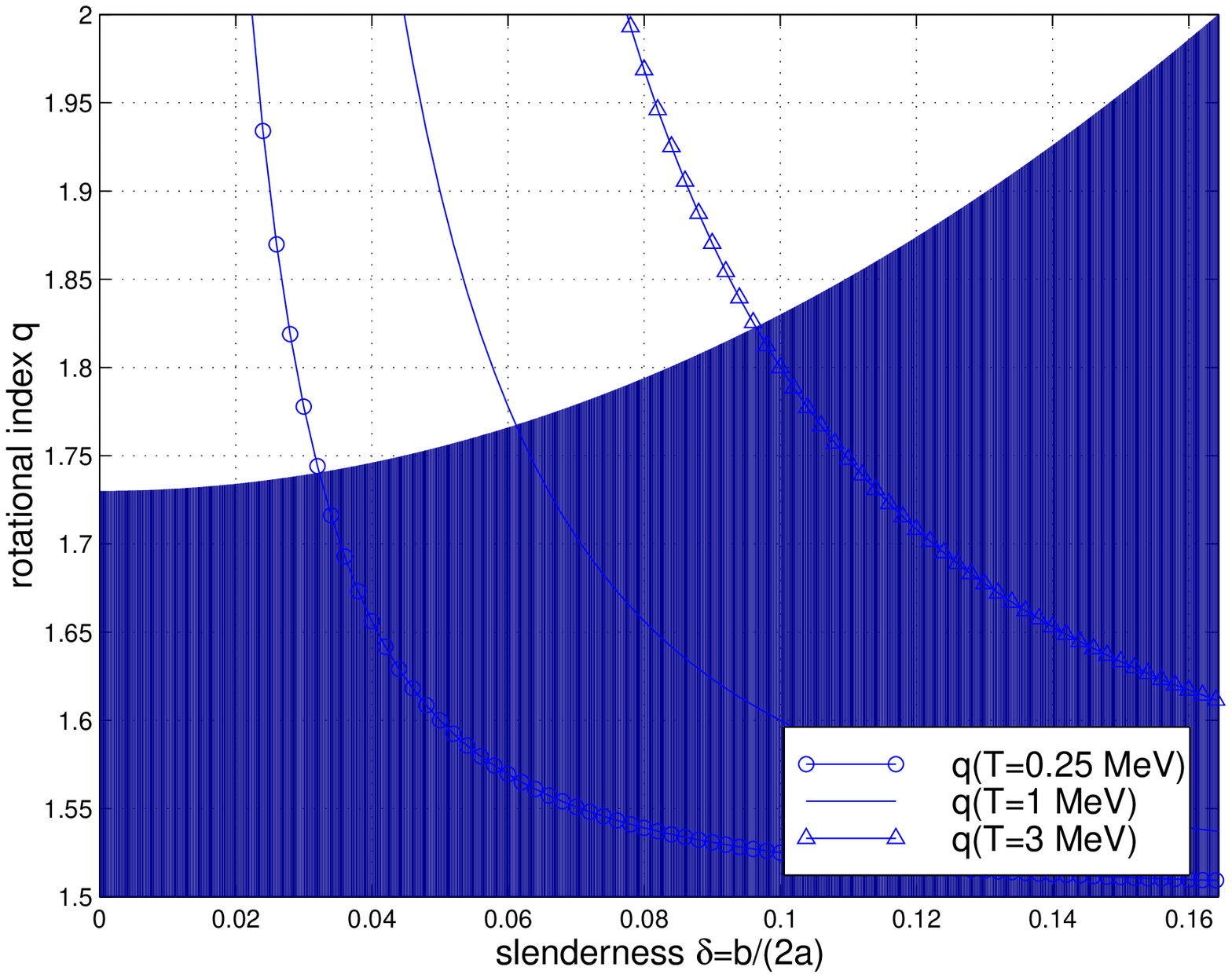}\includegraphics[angle=00,scale=.25]{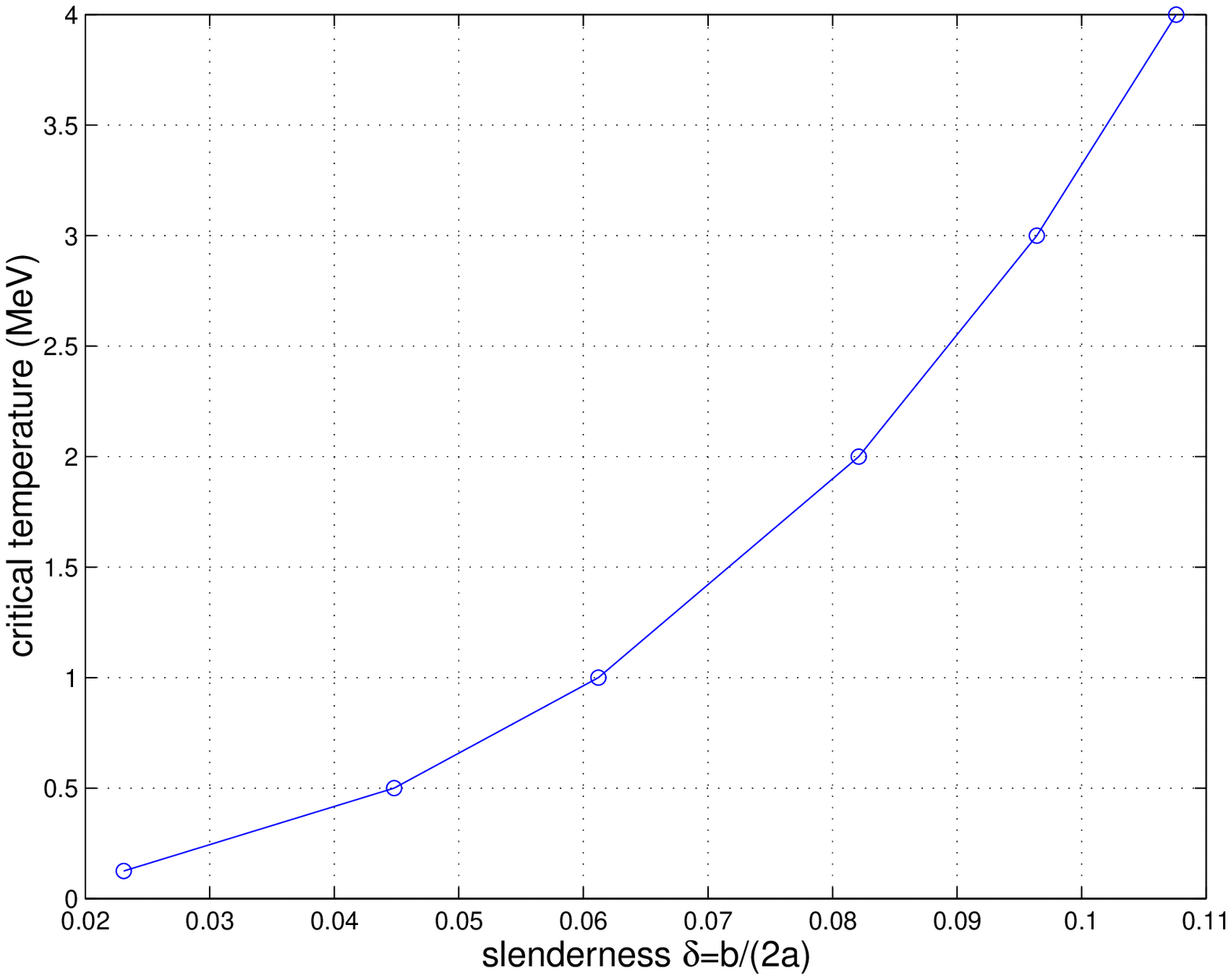}}
\caption{$(Left)$. Stability curve of a hot torus with finite slenderness
$\delta$ in terms of the rotational index $\sqrt{3}<q<2$, where $\sqrt{3}$
denotes the bifurcation point of all symmetry breaking modes 
in an infinitely slender torus and $2$ refers to the Rayleigh stability criterion 
for $m=0$. The results can be combined with a correlation of $q$ with pressure, here 
attributed to thermal pressure, though in practice augmented with magnetic pressure. The
result is a stability diagram showing critical temperature as a function of slenderness
$(right)$. The MeV temperatures involved are in good agreement with the anticipated
luminosities, powered by the central Kerr black hole, and dissipation thereof in 
suspended accretion.}
\end{figure}
\label{FIG_1201}
\end{center}

At the scale of 2 MeV we note comparable Alfv\'en and sound wave velocities,
\begin{eqnarray}
v_A\simeq 0.1 c,~~c_s\simeq 0.05 c.
\label{EQN_SA3}
\end{eqnarray}
Therefore, the Papaloizou-Pringle instability is generally facilitated by the combined 
contributions from thermal and magnetic pressures. After onset, it can be strengthed by 
the destabilizing effect of gravitational-radiation backreaction \cite{van02}. The 
latter introduces hysteresis: once the instability sets in, it is likely to persist.

Thus, sufficiently slender tori around rotating black holes have critical points
of thermal instability to the formation of non-axisymmetries. Similar results hold true 
in response to magnetic pressures. Whenever a slender torus forms, thermal and magnetic 
pressure-driven instabilities may have broad applications to tori at MeV as well as 
X-ray temperatures in light of (\ref{EQN_SA1},\ref{EQN_SA2},\ref{EQN_SA3}).

\subsection{Estimate of gravitational wave luminosities during spin down}

The luminosity in gravitational waves is determined by the nonlinear saturation 
amplitude for unstable non-axisymmetric modes in the torus in the presence of forced
MHD turbulence. Equipartition in a flat infrared spectrum in MHD turbulence at the 
threshold of stability (\ref{EQN_BSTAB}) predicts low-order mass inhomogeneities 
$\delta_m$ satisfy ${\delta m}/{M_T}\simeq {{\cal E}_B}/{{\cal E}_k}\simeq {1}/{15}$.
For tori (\ref{EQN_M1}) around supermassive black hole, the estimated luminosity in 
quadrupole emissions then satisfies
\begin{eqnarray}
L_{GW} = \frac{32c^5}{5G} \left(\frac{M}{R_T}\right)^5 \left(\frac{\delta M_2}{M}\right)^2
=7\times 10^{41} \left(\frac{R_T}{6M}\right)^{3} \left(\frac{M_{9}}{T_{7}}\right)^2\mbox{~erg s}^{-1},
\end{eqnarray}
at frequencies $f_{gw}=5 M_{6}^{-1} ({6.2}[(R_T/M)^{3/2}+(a/M)]^{-1})~\mbox{mHz}.$
This frequency is in the band width of sensitivity of the planned LISA mission. 
For emissions over a fiducial bandwidth $B=0.1\times B_{0.1}$ from a source at distance $D$, 
integration of $L_{GW}$ over a time $t_1$ in units of years gives a characteristic strain 
amplitude 
\begin{eqnarray}
h_{char} = \frac{\sqrt{2}}{\pi D} \sqrt{\frac{dE}{df}} 
= 1 \times 10^{-21} D_{100}^{-1}B_{0.1}^{-1/2} \left(\frac{R_T}{3M_{9}}\right)^{3/2}
   \left(\frac{M_{9}}{T_7}\right) t_{1}^{1/2}
\end{eqnarray}
where we ignored a redshift factor $1+z$ for the nearby source AGN of interest. 
\begin{figure}
\centerline{
\includegraphics[angle=00,scale=.30]{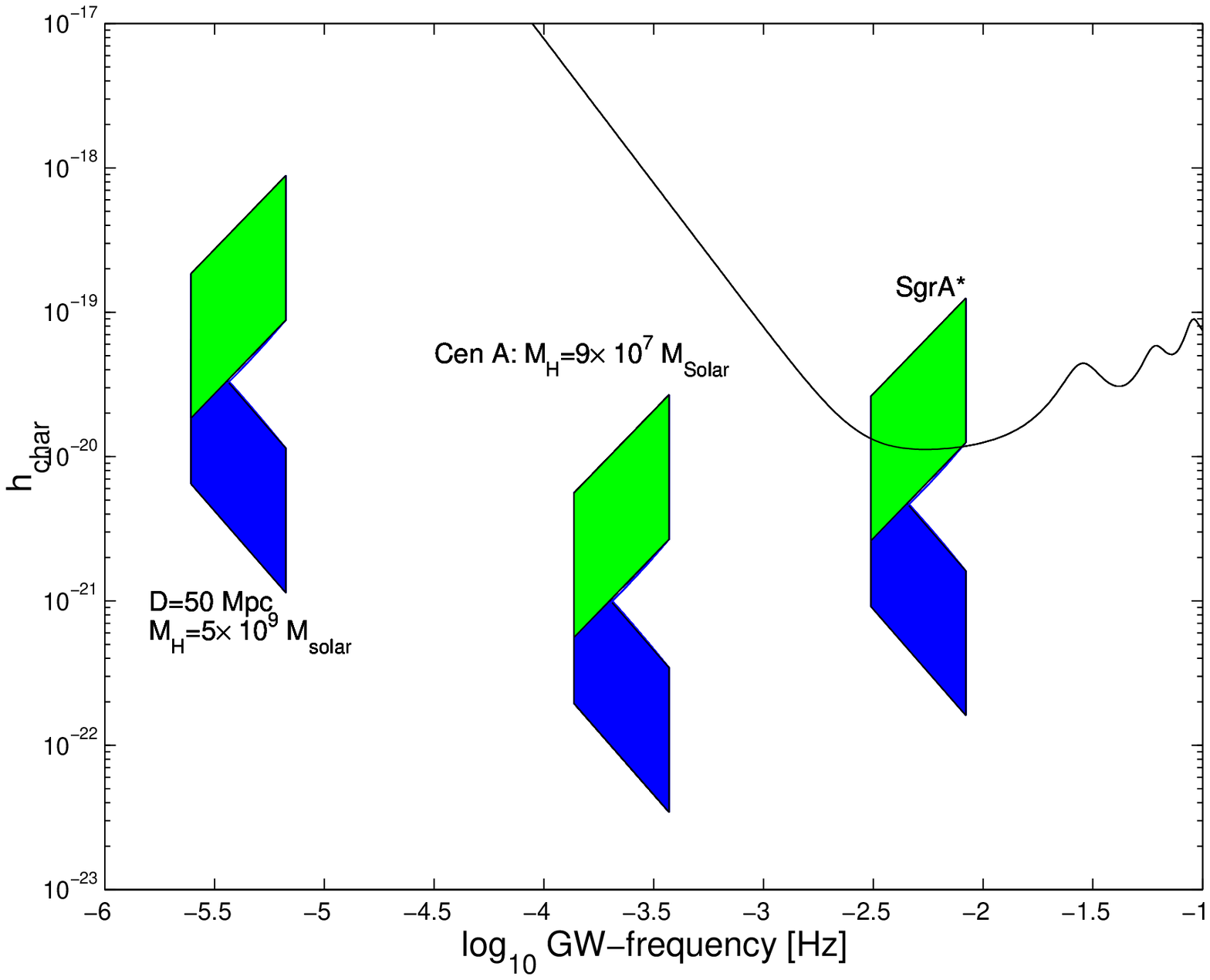}
\includegraphics[angle=00,scale=.30]{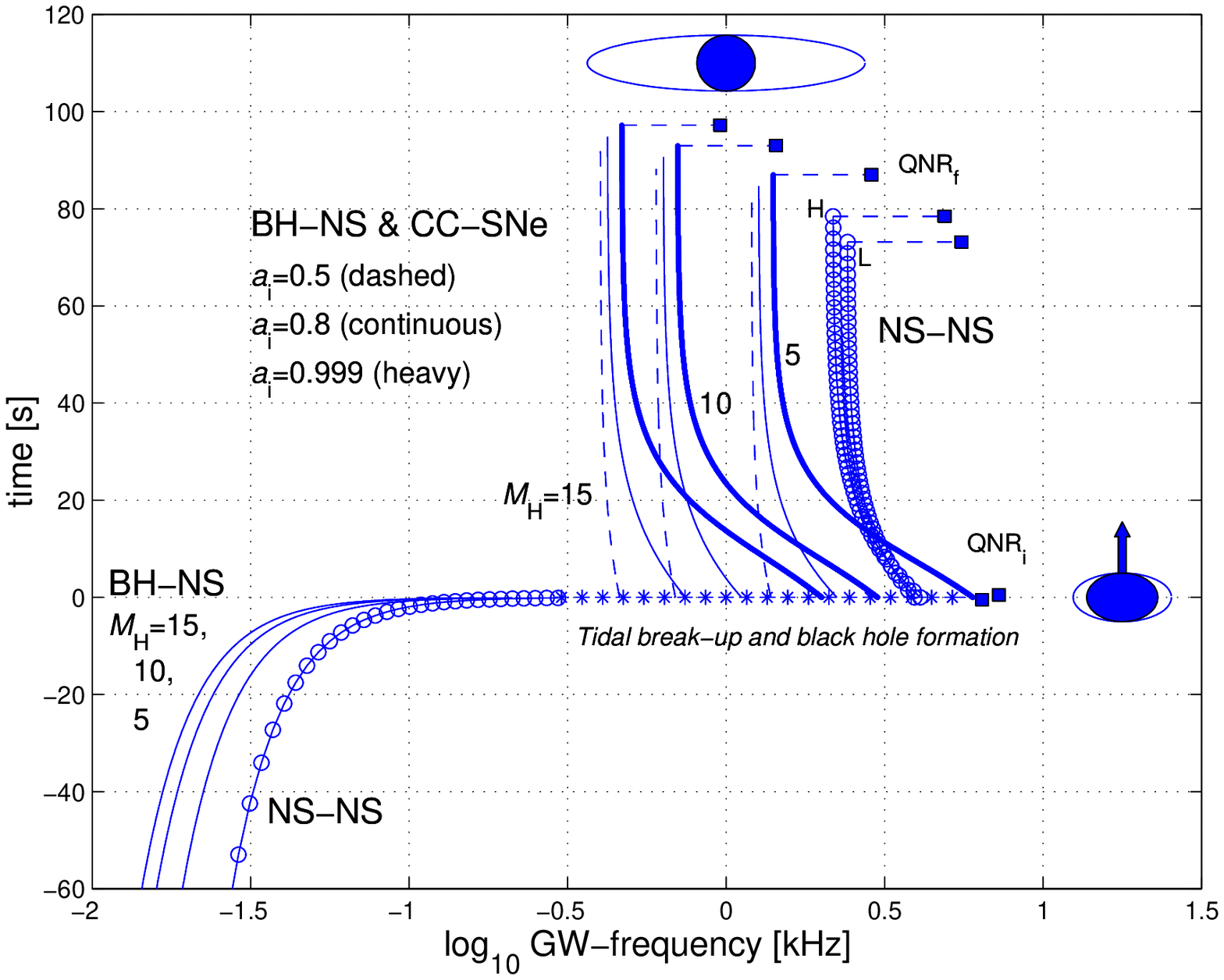}}
\caption{($Left$.) Line-emissions in gravitational waves around supermassive black 
holes compared with strain amplitude noise of LISA (continuous line) for a one-year 
integration time, assuming lifetimes of spin of $10^6$ yr and $10^7$ yr. The blue 
(downward) and green (upward) parallelograms refer co-evolution and constant disk 
mass, respectively. Only SgrA* appears to be of potential interest. 
($Right.$) Predicted chirps in gravitational radiation by long GRBs:
positive by mergers, negative in post-coalescence or in collapsars, where the 
naked inner engines in the former produce radio bursts for tens of seconds.
({Reprinted from} \cite{van09b}. (c)2009 The Royal Astronomical Society.)
\label{fs2}}
\end{figure}
For tori (\ref{EQN_M2}) around stellar mass black holes, the estimated gravitational
wave emissions are withing the frequency band width of sensitivity of Advanced 
LIGO-Virgo. The light curve in gravitational radiation features a {\em negative chirp} 
associated with the expansion of the ISCO during relaxation of a Kerr spacetime to a 
nearly Schwarzschild spacetime. The late-time gravitational wave frequency provides 
novel metrology of the mass of the black hole \cite{van08b}
\begin{eqnarray}
M \simeq 5.9~\left(\frac{f_{GW}}{1\mbox{~kHz}}\right)^{-1}M_\odot.
\label{EQN_M}
\end{eqnarray}
Based on \cite{bai08}, the frequency is about 2 k Hz at the end of spin down of
the Kerr black hole produced by the merger of neutron star binaries, and lower
by mergers of neutron stars with an initially rapidly rotating black hole 
companion. Frequencies of the latter may be very similar to those produced in
core-collapse events.  Fig. (\ref{fs2}) summarizes this outlook for LISA and LIGO-Virgo.

\section{X-ray tails (XRT) from GRB remnants}

The {\em Swift} discovery of X-ray tails (XRTs) points to 
to continuing accretion onto the black hole remnant of the prompt GRB phase \cite{kum08}. 
Similar accretion can occur from fallback of matter in mergers, post-GRB, as a result of 
messy break-up of the neutron star in the tidal field of a companion black hole 
\cite{lee98,lee99} or in the merger of two neutron stars \cite{ros07}. This can apply 
to GRB060614 when produced by an initially rapidly rotating black hole \cite{van09}.

Here, we attribute the duration of the prompt GRB emissions to the initial condition
on black hole spin, whereby fast (slow) initial spin produces long (short) GRBs. The
remnant at the end of the prompt GRB phase is a state at or about the fixed point
\begin{eqnarray}
\Omega_H=\Omega_T=\Omega_{ISCO}
\label{EQN_FXP}
\end{eqnarray}
of the system (\ref{EQN_EV1}) in model (\ref{EQN_MA}). It defines a {\em common state of
the inner engine to XRTs for all GRBs} with no memory of the initial spin rate of the
black hole. This restricts the ``universal central inner engine" recently hypothesized 
\cite{eic09} to XRTs, and does not include the preceeding GRB phase of the inner engine. 
Conceivably, (\ref{EQN_FXP}) defines the horizontal plateau's observed in some XRTs, as 
in GRB060614. The efficiency of converting matter accreting from the ISCO onto the black 
hole into radiation output is 25.34\% on the basis of (\ref{EQN_1st}). If most of it is 
released in X-rays, then
\begin{eqnarray}
L_X=0.25 \dot{m}
\label{EQN_LXM}
\end{eqnarray}
for an accretion rate $\dot{m}$. Any accompanying gravitational wave emissions from the
inner disk in this process are expected to be extremely weak, if present at all, with 
negligible increase in black hole mass and commensurate decrease in frequency 
in view of the observed X-ray luminosities, e.g., $L_X\simeq 10^{41}$ erg s$^{-1}$ in 
GRB060614 \cite{man07}. If unsteady, (\ref{EQN_FXP}) might account for large amplitude 
flaring as in GRB050502B \cite{geh09} by exchange of torques (\ref{EQN_EV1}) between
the black hole and the inner disk or torus of either sign, but likewise with no
noticeable change in spin of the black hole (e.g. \cite{eik03,lei08}).

\section{Is quantum gravity seen in the delay time in GRB080916C?}

GRB08016C reveals a striking delay time $\tau$ with a conservative upper limit of 16.5 s in 
the peak luminosity in its highest energy emissions \cite{taj09}. With $T_{90}=66$ s 
the normalized time delay satisfies
\begin{eqnarray}
\frac{\tau}{T_{90}}<25\%. 
\label{EQN_tau2}
\end{eqnarray}
Because of its large redshift $z=4.25$, the delay time in GRB080916C has received 
notable attention for its potential to provide novel constraints on or indications of 
quantum effects in the propagation of high energy photons. Violation of Lorentz 
invariance at high energies relative to a Planck-like quantum gravity mass scale 
$M_{QG}$ would give rise to a modified dispersion relation, resulting in an energy
dependent delay time \cite{ame98,ame09,eli09}
\begin{eqnarray}
\tau=\frac{E}{M_{QG}}H_0^{-1}\int_0^z\frac{(1+z)dz}{\sqrt{\Omega_\Lambda+\Omega_{m}(1+z)^3}},
\label{EQN_QG}
\end{eqnarray}
where $E$ denotes the energy of the photon and $H_0$ denotes the Hubble constant in a 
$\Lambda-$CDM cosmology with $\Omega_\Lambda=0.73$ and $\Omega_{m}=0.27$.

If all gamma-ray emission in GRB080916C comes from one component associated with the prompt 
GRB emission phase, then the match to (\ref{EQN_tau2}) suggests that $\tau$ is an intermediate 
time scale of a long-lived GRB inner engine, obviating the need for 
a modified dispersion relation of the type (\ref{EQN_QG}). The delay of 25\%
in (\ref{EQN_tau}) of model A (\ref{EQN_MA}) is a maximum for the observed delay time in 
neglecting photon background statistics. The apparent delay will be less in the presence
of a finite background photon count, whereby (\ref{EQN_tau}) is consistent (\ref{EQN_tau2}).
More likely, the highest energy gamma-ray emissions in GRB080916C represent secondary
emissions different from the prompt GRB phase with a broad range of possible 
delays $\tau$.

\section{Summary and outlook on multi-messenger surveys}

Kerr black holes are very similar to spinning tops in regards to their
energy and angular momentum (\ref{EQN_TOP}). An additional
frame dragging in their surroundings induces a powerful spin orbit coupling 
(\ref{EQN_E1}) and an interaction (\ref{EQN_JM}) with a surrounding torus 
magnetosphere.
 
The first provides a universal mechanism for creating high energy non-thermal 
emissions in capillary outflows. For black holes that are initially rapidly 
spinning, these emissions represent {\em minor} emissions in total energy output.
The second gives rise to viscous spin down of rapidly rotating black holes against
surrounding matter, thereby introducing a new secular timescale, the lifetime of 
black hole spin, in the process of emitting a {\em major} output in lower energy 
emissions by catalytic conversion of black hole spin energy.

Upstream of the terminal Alfv\'en front of a capillary outflow forms a linear 
accelerator for ionic contaminants in the funnel of ion tori in low-luminosity 
intermittent AGN, while at the same time downstream high energy photons can be 
produced in shocks \cite{van09}. 
In the process of spinning down the black hole, the torus is 
pressurized in forced MHD turbulence, and susceptable to non-axisymmetric 
instabilities. This promotes efficient shedding of angular momentum in gravitational 
waves.

For (\ref{EQN_E1}), model results (\ref{EQN_CUH1}-\ref{EQN_CUH2}) on UHECRs are 
consistent with statistical results Fig. \ref{FIG_ZAW} of the PAO. Model light 
curves by (\ref{EQN_E1}) and (\ref{EQN_EV1}) with closure (\ref{EQN_MA}) are in
excellent agreement with data BATSE, HETE II and {\em Swift} shown in Figs. 
(\ref{FIG_C1}) and (\ref{FIG_nLC}). 

The astrophysical consequences of the frame dragging induced interactions
of \S5 give the following answers to the challenges of \S2.
\begin{enumerate}
\item 
UHECR can be produced by low-luminosity, intermittent AGN at about GZK energies 
in the form of ionic contaminants in the funnel of a circumnuclear ion torus,
by linear acceleration upstream of a transient Alfv\'en front \cite{van09};
\item
Rotating black holes are universal inner ingines to GRBs in
core-collapse supernovae and mergers \cite{van01}, by dissipation \cite{mes06} 
in relativistic capillary jets downstream of a terminal Alfv\'en front \cite{van08a,van09}. 
Here, mergers produce long or short GRBs depending on the initial rate of spin 
of the black hole \cite{van01}.
The long duration event GRB060614 with no supernova is identified with a 
rotating black hole involved in the merger of a neutron star with a 
rapidly rotating companion black hole or another neutron star,
whose duration $T_{90}=102$ s represents the viscous lifetime of black hole 
spin \cite{van99,van01,van08b}. The same may apply to GRB050911 \cite{pag06} 
with no X-ray afterglow \cite{van09};
\item
The relativistic capillary jets are produced from first principles by gravitational
spin-orbit coupling (\S5.1) to (mostly) leptons along an open magnetic
flux tube supported by Carter's magnetic momemt in equilibrium. Powered by
black hole spin in a state of suspended accretion, super-Eddington luminosities 
are produced consistent with GRB data for magnetic fields satisfying (\ref{EQN_BSTAB});
\item
Rapidly rotating black holes act to pressurize surrounding matter into a torus,
which opens new radiation channels by catalytic conversion of black hole 
spin energy and angular momentum into a variety of emission channels \cite{van03}. 
In core-collapse events, magnetic torus winds provide a novel agent to 
power aspherical supernovae \cite{van03};
\item
Remants of long and short GRBs are slowly rotating black holes whose spin
rates have settled down to about the fixed point (\ref{EQN_FXP}) with no 
memory to their initial spin. It defines a common starting point for XRTs 
to all GRBs in the presence of continuing accretion of the outer envelopes 
of a progenitor star \cite{kum08} or remnant matter from a merger 
\cite{lee98,lee99,ros07,van09}.
\end{enumerate}
We note that after normalization, the delayed onset of peak luminosities at 
the highest energies in GRB080916C is less than 1, and is consistent with 
(\ref{EQN_tau2}) in the light curve produced by an initially maximally 
rotating black hole. This bound may be tested by future Fermi/{\em GLAST} 
detections of GRB080916C type events. Similar delays may be anticipated if, 
instead, the highest energy GRB photons represent secondary emissions.

We conclude that long GRBs are spin powered on 
the basis of the long duration merger event GRB060614, the spectral-energy
correlation and the matched filtering results shown in Figs. \ref{FIG_C1} and
\ref{FIG_nLC}, the requirement for a direct connection between the GRB emissions 
and the event horizon of a black hole\cite{lev93}, and super-Eddington
luminosities consistent with the true energies in gamma-rays in long GRBs \cite{van03}.

In light of the general model considerations of \S4, 
GRBs produced by frame dragging around rotating black holes may be compared 
with other models for inner engines to long GRBs.
Models of GRBs powered by hyperaccretion onto a newborn black hole \cite{kum08}
or a newborn neutron star \cite{dai06} derive long durations from the free fall 
timescale of matter from the progenitor stellar envelope. These models are not universal, 
in that hyperaccretion does not account for long duration time scales in mergers such as
GRB060614 \cite{del06} and, possibly, GRB050911 \cite{pag06}. Furthermore, accretion 
stimulates spin up \cite{kum08} or preserves the initial spin. This state of
the inner engine is difficult to reconcile with our observed decrease in the
normalized light curve shown in Fig. \ref{FIG_nLC}, when adhering to the connection
of the GRB emissions directly to the central object (as opposed to the accretion
disk). We note that the decrease in the normalized light curve produced in matched 
filtering is robust against the choice of template, in that it persists also for 
a block-type template \cite{van09}.

Models of GRBs powered by exploding dyadospheres (macroscopic regions with 
initially superstrong electric fields) interacting with a surrounding shell of 
matter \cite{ruf00} similarly do not apply to mergers and, furthermore, have no 
known scenario for their formation \cite{pag06a}. 

Emerging multi-messenger surveys promise to transform our view on the Transient Universe,
and probe their inner engines in novel ways. We mention just a few discovery 
opportunities implied by frame dragging around Kerr black holes:
\begin{itemize}
\item Repeat events from candidate UHECR-AGN associations, such as low luminosity 
      intermittent Seyfert galaxies, that may be accompanied by TeV emissions in
	  some cases. See \cite{far08} for related prospects;
\item QPOs in the electromagnetic spectrum from pressure-driven instabilities in tori 
      around rapidly rotating supermassive black holes, that may be amenable to
	  X-ray spectroscopy as in Fig. \ref{FIG_MCG} or radio observations (e.g. \cite{lac09}), 
\item Extragalactic radio bursts of short \cite{lor07} and long durations (tens of seconds)
      from naked inner engines to GRBs produced by mergers are of direct interest to, e.g.,
	  the Nan{\c c}ay Radio Telescope. Suitable all sky radio 
	  surveys may be provided by LOFAR (around 100 MHz) and Planck (30-857 GHz);
\item Low frequency QPOs in gravitational waves from SgrA*, whose luminosity is 
      conceivably at the threshold of detection of LISA;
\item Long bursts in gravitational radiation from long GRBs, from collapsars and
      mergers, featuring a negative chirp in the sensitivity range of Advanced
      LIGO-Virgo.	  
\end{itemize}

For multi-messenger surveys including LIGO-Virgo, a comprehensive parameter exploration of 
the existing LIGO-Virgo data-analysis pipelines appears to be relevant, notably by including 
searches for both positive and negative chirps combined. Joint surveys can be optimized by focusing on the 
local super-clusters \cite{ein94}, which contain about 1 million galaxies within 100 Mpc. This 
may be augmented by planned megaton-neutrino detectors for triggers of supernovae \cite{shi05}. 

\begin{acknowledgement}
The author gratefully acknowledges numerous stimulating discussions with Alessandro 
Spallicci, Alok C. Gupta, Giles Theureau, Isma\"el Cognard, Huub Rottgering, Robert 
S. Antonucci, Oliver Jennrich, 
Gerard 't Hooft, Alan J. Weinstein, and constructive comments from Amir Levinson. 
We thank Fabian Sch\"usser for kindly providing the data of Fig. \ref{FIG_A1}.
\end{acknowledgement}

\section{Appendix: Alfv\'en waves in capillary jets}

Alfv\'en waves are unique to MHD flows. The approximation of {\em ideal MHD} assumes 
negligible dissipation of the electromagnetic field in the fluid, corresponding to an 
infinite magnetic Reynolds number. It is valid for most large scale flows in astrophysics, 
such as extragalactic radio jets \cite{fan74}. If $p_B=B^2/8\pi$ and $e_B=B^2/8/pi$ denote 
the magnetic pressure and energy density in a magnetic flux tube of radius $R$, then the 
dissipationless limit implies adiabaticity when compressing the flux tube radially: 
$p_B(2\pi RdR)=d(\pi e_BR^2)$, i.e., the magnetic flux $\Phi=\pi BR^2$ is {\em frozen} 
into the fluid. Alternatively, a torsional perturbation applied to a flux tube creates 
an {\em Alfv\'en wave} with velocity \cite{lic67}
\begin{eqnarray}
v_A=\frac{B}{\sqrt{4\pi\rho+B^2}},
\label{EQN_VA}
\end{eqnarray}
where $\rho$ denotes the fluid density as seen in the comoving frame. The Alfv\'en wave
is rotational, exchanging angular momentum in the electromagnetic field and the fluid
while leaving density (and magnetic flux) invariant. Neglecting inertia, the Alfv\'en 
velocity reaches the velocity of light. 

If, furthermore, Reynolds stresses are neglected,
\begin{eqnarray}
F_{ab}j^b=0,
\label{EQN_FF}
\end{eqnarray}
we arrive at what appears to be first-order approximation to the magnetosphere surrounding 
rotating black holes \cite{bla77}. Note that (\ref{EQN_FF}) reduces the number of degrees of freedom 
in the electromagnetic field to two. For an electric current $j^b=\rho_e v^b$ produced by a 
charge density $\rho_e$ with four-velocity $v^b$, (\ref{EQN_FF}) implies $v^i\partial_iA_\phi=0$ 
and $v^i\partial_iA_0=0$ for a time-independent tube of flux surfaces $A_\phi=$const. along
the polar axis $\theta=0$. Thus, the electric potential satisfies $A_0=A_0(A_\phi)$,
and the electric field $\partial_iA_0=A_0^\prime\partial_iA_\phi$ as seen in the 
Boyer-Lindquist frame of reference is normal to the flux surfaces, whereby force-free flux 
surfaces are equipotential surfaces (\cite{gol69,bla77,tho86}).

The Alfv\'en front is a transition to equipotential surfaces downstream. It hereby
can communicate the raw frame dragging induced Faraday-induced potential (\ref{EQN_E1}) 
in the limit of $\omega=\Omega_H$ out to large distances powering a 
{\em linear accelerator upstream,} where the flux tube remains largely charge 
free except for ionic contaminants by UV-irradiation from a surrounding torus.

\end{document}